\documentstyle[aps,eqsecnum,epsfig]{revtex}

\begin{document}

\draft

\title{Subleading long-range interactions and violations of finite 
size scaling}

\author{Daniel Dantchev}

\address{Institute of Mechanics, Bulgarian Academy of Sciences, Acad.  
G. Bonchev St.  Building 4, 1113 Sofia, Bulgaria}

\author{Joseph Rudnick}

\address{Department of Physics, UCLA, Los Angeles, California 
90095-1547}

 \date{\today}
 
 \maketitle
\begin{abstract}
We study the behavior of systems in which the interaction contains a 
long-range component that does not dominate the critical behavior.  
Such a component is exemplified by the van der Waals force between 
molecules in a simple liquid-vapor system.  In the context of the mean 
spherical model with periodic boundary conditions we are able to 
identify, for temperatures close above $T_c$, finite-size 
contributions due to the {\it subleading} term in the interaction that are 
{\it dominant} in this region decaying algebraically as a function of $L$.
This mechanism goes beyond the standard formulation of the finite-size
scaling but is to be expected in real physical systems.
We also discuss other ways in which critical point behavior is 
modified that are of relevance for analysis of Monte Carlo simulations 
of such systems.

\end{abstract}


\section{Introduction}
\label{sec:intro}

An item of conventional wisdom in the study of critical phenomena is 
that the critical point behavior, including finite size scaling, is 
controlled by a relatively small number of features of a system, among 
which are the structure of the order parameter, the nature of boundary 
conditions, and the general properties of the interaction coupling the 
order parameter at different locations.  In particular, it is believed 
that short-range interactions lead to universal critical phenomena in 
the case of a given system.  For a non-critical $O(n)$ system with 
periodic boundary conditions, finite size corrections are expected to 
be exponentially small in the ratio $L/\xi$, where $L$ is the smallest 
of the system's linear dimensions, and $\xi$ is the correlation 
length.  This expectation holds everywhere on the phase diagram, with 
the possible exception of the coexistence curve, where, in certain 
systems, gapless spin wave excitations give rise to long-range 
correlations.

When interactions are long ranged, the above expectation is subject to 
revision.  The hallmark of a long-range interaction in the context of 
critical behavior is a diverging $n^{\rm th}$ moment.  That is, if 
$V(\vec{r})$ is long-range, then, for some sufficiently high $n$, the 
integral
\begin{equation}
V_{n} = \int V(\vec{r}) r^{n}d^{d}r
        \label{nmoment}
\end{equation}
diverges.  This diverging moment appears in the Fourier transform of 
the interaction, $v(\vec{q})$ through an anomaly in its expansion as a 
power series in $q$.  In the case of a very short range interaction, 
the power series expansion is entirely in integer powers of $q^{2}$.  
Any deviation from such an expansion represents an anomaly.  

That long-range interactions can alter the scaling behavior of a 
critical system has been known for some time \cite{FMN,S73}.  The 
first investigation of this phenomenon in the context of the 
renormalization group \cite{FMN} established that when $v(\vec{q}) 
\propto q^{\sigma}$ for sufficiently small $q$, with $\sigma <2$, then 
the critical point behavior of an interacting spin system differs 
fundamentally from that of a system in which the interaction is 
short-ranged.  The upper critical dimension $d_u$ for any such $O(n)$ 
system turns out to be $d_u=2\sigma$.  This has been established by 
renormalization group arguments in \cite{FMN} and rigorously proven in 
\cite{AF88}.  On the other hand, the lower critical dimension is 
$d_l=\sigma$ \cite{BZG76,FS82}.

In the context of critical phenomena, the criterion for short range 
interactions, with respect to the leading critical behavior, is a 
finite second moment of $V(\vec{r})$.  In terms of the power series 
expansion of $v(\vec{q})$, this means that whatever anomaly exists, 
does not interfere with, or dominate at small $q$, the first two terms 
in the expansion in powers of $q^{2}$.  That is, one can write for 
small $q$
\begin{equation}
v(\vec{q}) = v_{0} + v_{2}q^{2} + R(\vec{q})
        \label{expand1}
\end{equation}
where $R(\vec{q})$ is asymptotically smaller than the first two terms 
on the right hand side of (\ref{expand1}) for small $q$.  When the 
interaction between the order parameter at different points in the 
system has a Fourier transform that looks like the right hand side of 
(\ref{expand1}), one expects that the thermodynamic critical behavior 
will be as predicted by standard approaches for a system whose 
interaction is entirely short-ranged \cite{FMN}.  For finite-size 
$O(n)$ systems with $2<d<4$ this implies 
\cite{F72,FB72,B83,P90,BDT00}, among the other things, that the only 
relevant variable on which the properties of the finite system depend 
in the neighborhood of its bulk critical temperature $T_c$ is $L/\xi$.  
For $T>T_c$, finite-size corrections for systems having periodic 
boundary conditions are expected to be exponentially small in terms of 
$L/\xi$.  One assumes that for periodic boundary conditions all 
reference lengths, aside from the bulk correlation length $\xi$, will 
lead only to corrections to the above picture.

The above expectation has been challenged in recent papers by Chen and 
Dohm \cite{C&D1,C&D2,C&D3,C&D4,C&D5}.  In these papers, it is pointed out 
that a model having both short-range interactions and a sharp, 
wavelength-dependent cutoff of fluctuations $\Lambda$, will exhibit 
finite size 
corrections to the infinite system limit that swamp those 
traditionally associated with finite size scaling.  They demonstrate 
that for such a particular choice of the fluctuation cutoff function, 
one will observe corrections to the infinite system thermodynamic 
behavior going as an inverse power law in $L$ that do depend also on
$L\Lambda$ and not only on $L/\xi$.

A careful investigation of the model discussed by Chen and Dohm 
reveals that the power law contributions to the finite size 
corrections result from the interplay of two features of that model.  
The first is a sharp cutoff of fluctuations in momentum space and the 
second is the removal of the ``remainder'' term in (\ref{expand1}), 
which has the effect of introducing an interaction that cannot be 
periodic in reciprocal space.  The combined effect of these two 
features is an effective interaction that falls off as a power law in 
the separation between degrees of freedom.  This power-law interaction 
leads immediately to power-law contributions to the finite size 
corrections.  The last result provides a natural explanation for the 
discrepancies between the finite-size effects that are predicted by 
infinite cut-off field-theoretical schemes \cite{B82,BZ85,RGJ} and the 
finite size effects that arise in theories with a sharp cut-off 
\cite{C&D1,C&D2,C&D3,C&D4,C&D5}.  

That a sharp cutoff is essential to the appearance of power-law 
corrections to the infinite system limit was noted by Chen and Dohm in 
\cite{C&D2}, where an example of smooth cutoff effects is also 
presented.  The both cases of sharp and smooth cut-off are clearly 
distinguished in \cite{CD00} for systems with dimensionality
$d>4$. For $d>2$ they also realized a close relation between a
non-exponential large-distance behavior of the bulk correlation
function (due to the sharp cut-off) and the power-law finite size
behavior above $T_c$ \cite{C&D5}.   
Nevertheless, they do not consider in any of their articles the 
observed power-law finite-size corrections essentially as a result of 
an effective long-range interaction. 

In the present article we will deal only with the case when the 
hyperscaling holds, i.e.  we will suppose $2<d<4$, but definitely 
similar effects will be observed also above the upper critical 
dimension $d_u=4$, which case is a subject of an intensive discussion 
in the literature in the last time (see, e.g.  \cite{CD00,LBB99} and 
the literature cited therein).

It is certainly true that the degrees of freedom of a system of spins 
on a lattice is represented in reciprocal space by wave vectors 
confined to a Brillouin zone.  However, the interactions between the 
spins have a Fourier-transformed form that is periodic in the zone.  
That is, the momentum-space version of the interaction is not purely 
quadratic, but rather reproduces itself as the wave number is shifted 
by a reciprocal lattice vector.  Such interactions do not give rise to 
the non-scaling terms obtained by Chen and Dohm.  An alternative 
source of the fluctuation cutoff is a Fermi surface.  However, while 
the Fermi surface is a natural construct in the case of 
non-interacting Fermions, it is hard to imagine a property of any 
actual system that allows Bose excitations to be freely occupied in 
one regime of wave-vector space and that forbids the occupation of 
those excitations in an immediately adjoining neighborhood.  No 
specific scenario leading to this behavior has yet been proposed, at 
least to our knowledge.

To recapitulate, the two elements that are required to obtain the 
results of Chen and Dohm are, first, an interaction appropriate to a 
spin system in a continuum, and, second, a sharp cutoff in the 
fluctuation spectrum that mimics either the effect of a Fermi surface 
at $T=0$, or a Brillouin zone.  In Appendices \ref{app:cutoff}, 
\ref{app:truncation} and \ref{app:soft}, we explore the interplay of 
these two elements.  We demonstrate that neither one alone suffices to 
to give rise to the effective long-range interaction that leads to a 
violation of finite size scaling.  Appendix \ref{app:derivation} 
supplies details of the analysis of Chen and Dohm that are central to 
the derivation of their predictions in terms of violation of finite 
size scaling.  These details which are missing from their papers, are 
intended as an aid to those interested in a critical study of the 
basis for their results.  The authors of this work wish to note that 
they are in full agreement with the mathematical conclusions drawn by 
Chen and Dohm.

Regardless of the relevance of their particular model to either 
physical reality or the fictitious, but nevertheless physically 
meaningful, world of simulations, Chen and Dohm have raised an 
interesting point.  What can one reasonably expect to occur when 
interactions are intrinsically long-range, even if the range of the 
interactions is not sufficiently great to alter the asymptotic 
critical behavior of the system in question?  Will these 
``subleading'' long-range interactions inevitably alter the sorts of 
finite size effects that are naively expected to be present in finite 
systems with short-range interactions and periodic, or, indeed, 
arbitrary, boundary conditions?

In this paper, we address this question in the context of an 
interaction whose Fourier transform allows for the kind of small-$q$ 
expansion shown in Eq.  (\ref{expand1}).  The term $R(\vec{q})$, 
asymptotically smaller than the first two contributions to the right 
hand side of that equation when $q$ is small, contains a component 
going as $q^{\sigma}$, where $\sigma/2$ is noninteger, and $\sigma >2$.  
This means that the interaction is long range, but that
\begin{equation}
\int V(\vec{r})r^{p}d^{d}r
        \label{diverging}
\end{equation}
is finite for $p \le \sigma$.  Alternatively, we imagine a 
$V(\vec{r})$ going as $r^{-d-\sigma}$ for large values of $r$.  Such 
an interaction is far from unphysical.  In fact, the van der Waals 
interaction, which decays in three dimensions as $r^{-6}$, is 
consistent with $\sigma =3$.  The explicit calculations presented in 
the article are for the case $2<d<4$, $2<\sigma<4$ where $d+\sigma\le 
6$.  We find that an interaction of this form does, indeed, give rise 
to interesting modifications of the critical point behavior of a 
spin-like system.  Those modifications include, but are not limited 
to, {\it power-law finite system contributions that dominate the 
exponentially damped terms arising from a standard analysis of short 
range systems}.  In other words, in contrast to the infinite systems, 
where the {\it subleading} terms in the interaction are producing only 
{\it corrections} to the thermodynamic critical behavior, for finite 
systems they lead to {\it dominant} finite-size contributions in a 
given regime.  The case $\sigma = d = 3$ is special, in that there are 
logarithmic corrections to the nominally power law finite size 
corrections.

The current understanding of the consequence of power law interactions 
with regard to critical point properties of an $O(n)$-model spin 
system is fairly well-developed, but is not yet complete.  Assuming 
that the interaction is of the form $r^{-d - \sigma}$, then there are 
three regimes, depending on the magnitude of positive exponent 
$\sigma$.  First, if $\sigma <d/2$, then the critical behavior is 
mean-field-like.  On the other hand, if $\sigma >2$, the behavior of 
the system in the immediate vicinity of the critical point is as if 
the interactions are short-ranged.  Clearly, when $d\ge 4$, the two 
regions overlap, and the critical behavior of the system is always 
expected to be as predicted by mean field theory.  In fewer than four 
dimensions, the regime $d/2 < \sigma < 2$ represents an intermediate 
case, in which the critical behavior is not necessarily dominated by 
either the mean field or the renormalized short-range interaction 
limits.  This regime has been investigated for the $O(n)$ model in the 
context of a renormalization group-based expansion in $2 \sigma - d$ 
by Fisher, Ma and Nickel \cite{FMN}.  In the case of a one-dimensional 
Ising system, it has been argued that when $\sigma =1$, the critical 
properties are intimately related to those of the two-dimensional $XY$ 
model \cite{K1}, in that the appropriate version of the 
Renormalization Group is in the same generic class as the equations 
shown to apply to the latter model by Kosterlitz \cite{K2}. The
existence of a phase transition in this borderline case has been
rigorously proven by Fr\"{o}lich and Spencer \cite{FS82}. Later Imbrie
and Newman \cite{IN88} presented a rigorous proof of the existence in
the system of a phase in which the two-point correlation function
exhibits power-law decay with an exponent that varies continuously in
a finite temperature range below the critical temperature.

At this point in time, there seems to be no serious controversy on the 
analytical front.  However, in a recently published paper Bayong and 
Diep \cite{B&D1} have utilized Monte Carlo simulations to investigate 
the critical properties of an continuous Ising system (i.e.  the spin 
variables of the model can take any value between $-1$ and $1$) in one 
and two dimensions with an interaction of the form $1/r^{d+\sigma}$.  
They seek to determine whether the behavior of this system is 
consistent with predictions based on Renormalization Group methods and 
other analytical approaches.  While they are able to identify trends 
for the critical exponents that qualitatively follow the assertions of 
previous investigators, the values of the exponents are not in concert 
with those obtained in analytical investigations.  This calls into 
question either basic assumptions with respect to the influence of 
long-range interactions on critical point properties, or on the 
validity of Monte Carlo methods, at least as exploited by the authors 
(e.g.  for the case $d=1$, $\sigma=1.1$ rigorous mathematical proof 
can be presented \cite{R2000} for the absence of phase transition at 
finite temperature, while the authors determine the critical exponents 
around such a transition).  Alternatively, there is the possibility 
that more care and experience is needed in the way in which long range 
interactions are discussed analytically.  For example, will the 
long-range interactions studied by Bayong and Diep induce 
complications in the finite size corrections that require a more 
sophisticated approach than was undertaken by them?

Additionally worthy of mention are the results of Luijten \cite{L99} 
that indicate the existence of discrepancies between the 
Renormalization Group predictions for the behavior of the Binder 
cumulant $B$ (for a definition of $B$ see \cite{B81}) in a fully 
finite system with periodic boundary conditions at its bulk critical 
temperature and the numerical results for this quantity obtained by 
Monte Carlo simulations.  Here, the focus is a system with leading 
power-law interaction.  As a function of $\varepsilon=2\sigma-d$ 
Renormalization Group predicts \cite{L99,CT2000} that $B$ depends in a 
leading order on $\sqrt{\varepsilon}$, i.e.  in the same way as in the 
case of short-range systems \cite{BZ85}, while numerical results 
\cite{L99} suggest a linear dependence on $\varepsilon$.  The 
numerical simulations have been performed for a discrete Ising model 
with $d=1$ and $d=2$ and periodic boundary conditions.  It is 
difficult to comment on the origin of this discrepancy---one is 
tempted to suggest a more careful analysis of the numerical data, 
despite the fact that the cluster algorithm \cite{L95} used by Luijten 
is able to take into account the interaction of any spin with all the 
other spins in the system, including the infinite sequence of images 
under the periodic boundary conditions.  In other words, no truncation 
of the interaction has been enforced.

The outline of this paper is as follows.  In Section \ref{sec:model}, 
we outline the general features of the model to be studied: a mean 
spherical model in confined to a $d$-dimensional hypercube of length 
$L$ per side, subject to periodic boundary conditions.  The 
interaction contains a component with a diverging higher moment.  The 
signature of this component is a term in the Fourier transform of the 
interaction going as $q^{\sigma}$ with $2< \sigma < 4$.  As there is 
also a term that goes as $q^{2}$, the interaction is short-ranged, in 
that critical exponents are those associated with interactions having 
a finite second moment.  Section \ref{sec:eos} contains a detailed 
analysis of the equation of state of this model, special attention 
being paid to the influence of the long-range portion of the 
interaction.  It is found that an expansion in the strength of this 
contribution to the interactions between degrees of freedom in the 
system suffices to establish the key characteristics of the 
system---in particular, the finite size corrections to asymptotic 
critical behavior.  This section provides all the mathematical detail 
needed to extract both the asymptotic critical behavior and the 
leading order corrections arising from non-leading-order long range 
interactions.  In Section \ref{sec:chi} the results of Section 
\ref{sec:eos} are utilized to discuss the dependence of the isothermal 
susceptibility of this system on both reduced temperature, $t$ and 
system size, $L$ in the important regimes surrounding the critical 
point.  The case $d = \sigma = 3$ is given special treatment, as in 
this special instance, which is relevant to a system in which van der 
Waals interactions play a role in ordering.  Here, exponent 
``degeneracy'' gives rise to logarithmic corrections to pure power law 
behavior.

We find that one can easily envision situations in which corrections 
to scaling, in the form of contributions to the thermodynamics of a 
finite system that scale as $L$ to a non-leading power are of a 
magnitude comparable to the putative leading order terms.  This 
indicates that there are circumstances in which the analysis of 
simulation data must be undertaken with care.

\section{The model}
\label{sec:model}

In this paper, we restrict our attention to the fully finite mean 
spherical model.  We assume a $d$-dimensional hypercube of length $L$ 
per side with periodic boundary conditions.  The boundary conditions 
are consistent with the way in which one sets up a model for Monte 
Carlo investigations, in that the system is finite, but lacks physical 
boundaries.  The degrees of freedom consist of a set of $N$ localized 
spins with gaussian weight, and the Hamiltonian is given by
\begin{equation}
{\cal H} = -\sum_{i,j} V\left(\vec{R}_{i}-\vec{R}_{j} \right) 
s_{i}s_{j} - h \sum_{i}s_{i}
        \label{ham1}
\end{equation}
where $\vec{R}_{i}$ is the position vector of the $i^{\rm th}$ spin.  
The Fourier transform of the interaction, 
$V\left(\vec{R}_{i}-\vec{R}_{j}\right) \ge 0$ is assumed to possess the 
following low-$q$ expansion
\begin{equation}
v(\vec{q}) = v_{0} + v_{2}q^{2} +v_{\sigma} q^{\sigma}+v_4 q^4.
        \label{expand2}
\end{equation}
where $2 < \sigma < 4$, $v_0, v_{\sigma} > 0$, and $v_2, v_4 < 0$.  
The term $v_{\sigma} q^{\sigma}$ in (\ref{expand2}) is associated with 
a contribution to the real-space version of the interaction going as 
$|\vec{R}_{i} - \vec{R}_{j}|^{-d-\sigma}$, \underline{as long as 
$\sigma$ is not an even integer} (in the opposite case one will have 
in addition some logarithmic corrections).  Note that the signs of the 
coefficients in the small $q$ expansion are chosen so as they normally 
appear for interactions that decay in power laws with the distance 
between the interacting objects - molecules or spins (this easily can 
be checked, say, for the example of a one dimensional system with such 
an interaction; $v_0$, $v_2$, $v_\sigma$ and $v_4$ are 
$\sigma$-dependent---for simplicity of the notations this dependence 
is omitted here).  Furthermore, we suppose that $v(\vec{q})-v_0<0$ if 
$\vec{q}\ne \vec{0}$, which reflects the fact that there are no 
competing interactions in the system we consider and that the only 
ground state of this system is the ferromagnetic one.  Of course, it 
would be interesting to consider such systems---say a combination 
between antiferromagnetic short range and ferromagnet subleading 
long-range interactions, but this is out of the scope of the current 
article.

The partition function of this system is equal to the multiple 
integral
\begin{equation}
\int_{-\infty}^{\infty} ds_{i} \cdots \int_{-\infty}^{\infty}ds_{N} 
\exp \left[ -\beta {\cal H} \right]
        \label{partfun1}
\end{equation}
supplemented by the mean spherical condition
\begin{equation}
\sum_{i=1}^{N} \langle s_{i}^{2}\rangle =N
        \label{msc1}
\end{equation}
which can be enforced with the use of a ``Lagrange multiplier'' term 
going as $ \lambda \sum_{i=1}^{N}s_{i}^{2}$ into the effective 
Hamiltonian, and thence into the partition function.  The spherical 
model equation of state then takes the form
\begin{equation}
\sum_{\vec{q}}\frac{k_{B}T}{ \lambda-(v_{0} + v_{2}q^{2} + 
v_\sigma q^{\sigma}+ v_4 q^4) } = N
        \label{msc2}
\end{equation}
The phase transition in this model occurs when the combination 
$(v_{0}-\lambda)/v_2 \equiv r$ takes on a value asymptotically close 
to zero.  The difference between the equation of state in (\ref{msc2}) 
and the standard mean spherical model condition in short range systems 
lies in the addition of the term going as $q^{\sigma}$ in the 
denominator on the left hand side of (\ref{msc2}).  In general, this 
term is taken to be negligible, but we will soon see that it leads to 
interesting effects.

Because we are looking at a finite model, the sum over $\vec{q}$ in 
(\ref{msc2}) is subject to restrictions.  In particular, under the 
assumption of a periodically continued hypercubic system of length $L$ 
per side, allowed values of $\vec{q}$ are of the form $\vec{q} = 2 \pi 
\vec{n}/L$, where $\vec{n}$ is a vector with integer components.  A 
number of methods have been developed for the evaluation of the kind 
of sum in the equation of state (\ref{msc2}).  When there is no term 
going as $q^{\sigma}$, the sum is quite standard, and has been 
performed (neglecting the term proportional to $k^4$) with the use of 
a number of techniques, including the Poisson sum formula and variants 
on the Ewald summation trick \cite{ewald}.  The addition of the 
non-integral power of $q$ into the denominator in (\ref{msc2}) 
complicates matters a bit, but adaptations of the above methods to the 
case $0<\sigma<2$ have proven effective \cite{B&T,B89}.  Appendices 
\ref{app:alternative} and \ref{app:general} outline such adaptations 
that incorporate also the case $2<\sigma<4$.  They make use of contour 
integration tricks to ``map'' the summation onto the more conventional 
short-range one.  In this paper, we make use of analytical methods 
based on the relationship between expressions central to the 
statistical mechanics of this system and well-known mathematical 
functions.

\section{The equation of state}
\label{sec:eos}

The equation of state for the mean spherical model (for a 
comprehensive review on the results available for this model see 
\cite{BDT00}) is
\begin{equation}
K=\frac{h^2}{Kr^2}+W_{d}^{\sigma}(r,b,c,L|\Lambda),
\label{sfe}
\end{equation}
where $q=|{\bf q}|$, $K=|v_2|/(k_B T)$, $b=v_\sigma/v_2<0$, 
$c=v_4/v_2>0$, $h$ is a properly normalized external magnetic field, 
and
\begin{equation}
W_{d}^{\sigma}(r,b,c,L|\Lambda)=\frac{1}{L^d}\sum_{\bf 
q}\frac{1}{r+q^2+bq^\sigma+c q^4}.
\label{wdef}
\end{equation}
Here $\bf q$ is a vector with components $q_j=2\pi m_j/L$, $m_j=\pm 
1,\pm2, \cdots$, $j=1,2, \cdots, d $, in the range $-\Lambda\le 
q_j<\Lambda$.  The critical point of this system is  given by 
$K_c(d,\sigma,b,c,\Lambda)=W_{d}^{\sigma}(0,b,c|\Lambda)$, where 
$\Lambda=\pi/a$, $a$ being the lattice spacing for a lattice system 
(or $\Lambda$ is  the finite cutoff of the corresponding field theory), 
and
\begin{equation}
W_{d}^{\sigma}(r,b,c|\Lambda)=\frac{1}{(2\pi)^d}\int_{-\Lambda}^{\Lambda}d^dq 
\cdots \int_{-\Lambda}^{\Lambda}\frac{1}{r+q^2+bq^\sigma+cq^4}.
\label{bulk}
\end{equation}
Wherever possible we will omit the contributions that are due to the
term proportional to $q^4$. Because of that we will omit $c$ in the
remainder of the text in the symbols $W_{d}^{\sigma}(r,b,c,L|\Lambda)$
and $W_{d}^{\sigma}(r,b,c|\Lambda)$. The term proportional to $q^4$
is included in (\ref{wdef}) and (\ref{bulk}) in order only to
assure that no artificial poles will exist in the denominator of
$W_{d}^{\sigma}(r,b,L|\Lambda)$ and in the integrand of 
$W_{d}^{\sigma}(r,b|\Lambda)$ at large values of $q$ (we recall that
$b<0$ but $c>0$ and that the propositions we made for $v(\vec{q})$
guarantee that there is no real root of the equation $1+b
q^{\sigma-2}+c q^2=0$).

We are interested in the behavior of the finite system close to or 
below the critical temperature $K_c$, when $r$ in the right-hand side 
of equation (\ref{sfe}) is small (i.e.  when $0<r\ll 1$).  As is clear 
from (\ref{sfe}) and (\ref{bulk}) the singularities in the behavior of 
$W_{d}^{\sigma}(r,b|\Lambda)$ as a function of $r$, which in turn can 
be transformed as singularities with respect to the temperature, arise 
for small values of $q$.  In what follows we will retain only those 
contributions to the behavior of the quantities involved that are 
associated with the effects of long-range fluctuations (i.e.  $q\ll 
1$).  Proceeding in this way, we obtain
\begin{eqnarray}
W_{d}^{\sigma}(r,b,L|\Lambda) &\simeq &\frac{1}{rL^d}+ S_L(d,r,2|\Lambda)- 
\frac{b}{L^d}\sum_{\bf q}\frac{q^\sigma}{(r+q^2)^2} \nonumber \\
&=&\frac{1}{rL^d}+S_L(d,r,2|\Lambda)-b(1+r\frac{\partial}{\partial r}) 
S_L(d,r,\sigma|\Lambda),
\label{eosa}
\end{eqnarray}
where
\begin{equation}
S_L(d,r,\sigma|\Lambda)=\frac{1}{L^d}\sum_{{\bf q}\ne {\bf 
0}}\frac{q^{\sigma-2}}{r+q^2}.
\label{sdef}
\end{equation}
To analyze the finite-size behavior of $S_L(d,r,\sigma|\Lambda)$ we 
make use of the identity
\begin{equation}
\frac{q^{2p}}{(r+q^2)^a}=\int_{0}^{\infty}\exp(-q^2 
t)\frac{t^{a-p-1}}{\Gamma(a-p)} \ _1F_1(a;a-p;-rt)dt,
\label{fi}
\end{equation}
where $a>p$ and $_1F_1$ is the confluent hypergeometric function.  For 
$a=1$ and $p<1$ this identity further simplifies to
\begin{equation}
\frac{q^{2p}}{r+q^2}=\int_{0}^{\infty}\exp[-(q^2 +r)t] 
t^{-p}\gamma^*(-p,-rt) dt,
\label{si}
\end{equation}
where $\gamma^*(a,x)$ is a single-valued analytic function of $a$ and 
$x$, possessing no finite singularities \cite{AS}
\begin{equation}
\gamma^*(a,x)=e^{-x}\sum_{n=0}^{\infty}\frac{x^n}{\Gamma(a+n+1)}= 
\frac{1}{\Gamma(a)}\sum_{n=0}^{\infty}\frac{(-x)^n}{(a+n)n!}, \ \ 
|x|<\infty.
\label{gammastar}
\end{equation}
Both of the identities (\ref{fi}) and (\ref{si}) can be proven by 
integrating by parts the corresponding series representations of 
$_1F_1$ and $\gamma^*$. In \cite{KT91} the identity (\ref{fi}) has been
used to analyze the finite-size behavior of $O(n)$ model with both a leading 
long-range interaction of the type $1/r^{d+2-2\alpha}$, $\alpha\rightarrow 0$ 
and a short-range interaction present in the system.

With the help of this identity one obtains
\begin{eqnarray}
S_L(d,r,2(p+1)|\Lambda)&=&\int_{0}^{\infty}e^{-rt}t^{-p}\gamma^*(-p,-rt) 
\left[\frac{1}{L^d} \sum_{{\bf q}\ne {\bf 0}} e^{-q^2 
t}\right]dt \nonumber \\
&=& I_{{\rm bulk}}^p(r,d|\Lambda)+L^{2-d-2p}I_{{\rm 
scaling}}^p(rL^2,d)+ \nonumber \\
& & \left( \frac{\Lambda}{2\pi}\right)^{d+2p-2} M^{-2} I_{{\rm cut \ 
off}}^p(r/\Lambda^2),
\label{sresult}
\end{eqnarray}
where $M=L\Lambda/(2\pi)$, and
\begin{equation}
I_{{\rm bulk}}^p(r,d|\Lambda)=\int_0^\infty 
e^{-rt}t^{-p}\gamma^*(-p,-rt) \left[ \frac{{\rm erf}(2\pi \Lambda 
\sqrt t)}{\sqrt{4\pi t}} \right]^d dt,
\label{ib}
\end{equation}
\begin{equation}
I_{{\rm scaling}}^p(x,d)=\int_0^\infty e^{-xt}t^{-p}\gamma^*(-p,-xt) 
\left\{ \left[ \sum_{k=-\infty}^{\infty}e^{-4\pi^2 k^2 t} \right]^d - (4\pi 
t)^{-d/2}-1 \right\}dt,
\label{is}
\end{equation}
and
\begin{equation}
I_{{\rm cut \ off}}^p(x)=-\frac{4}{3}\pi^2 d \int_0^\infty 
e^{-4\pi^2(1+x)t}{\rm erf}(2\pi \sqrt t)t^{1-p}\gamma^*(-p,-xt) dt.
\label{ic}
\end{equation}
Taking into account the fact that $\gamma^*(p,x)\rightarrow 1$ when 
$p\rightarrow 0$ (see Eq.(\ref{gammastar})) it is clear that when 
$\sigma=2$ all these expressions give the corresponding well-known 
results (see, e.g.  \cite{C&D2}) with only short-range term in the 
interaction.  We will treat the bulk term separately.  To derive the 
behavior of this term, it is not necessary to use the representation 
given here.  In fact, it is relatively straightforward to derive the 
leading $r$ dependence of the bulk term due to the existence of a 
subleading long-range term in the interaction.

Furthermore, it is obvious that the term that contains a contribution 
due to the finite cut-off is not, in fact, correct, because the 
expansion we have utilized guarantees only that effects due to small $q$ 
behavior are properly taken into account.  In what follows we will 
simply ignore the precise form of this term.  The ``finite-size 
scaling term'', insofar as it is due to long-wavelength 
contribution, is calculated exactly.  We are able to conclude that the 
equation of state can be written in the form
\begin{eqnarray}
K &=& \frac{1}{rL^d}+\frac{h^2}{Kr^2}+W_{d}^{\sigma}(r,b|\Lambda)+ 
L^{2-d}I_{{\rm scaling}}^0(rL^2,d) \\
& & -bL^{4-d-\sigma}(1+r\frac{\partial}{\partial r}) I_{{\rm 
scaling}}^{(\sigma-2)/2}(rL^2,d)+ \Lambda{\rm-dependent \; term 
}\nonumber.  \\
\label{es}
\end{eqnarray}
In Appendix \ref{app:Lambdep} we show that the $\Lambda$-dependent 
terms can be neglected for the purposes of the analysis carried out 
here. For the bulk term $W_{d}^{\sigma}(r,b|\Lambda)$ for small $r$ it can 
be shown by using the standard techniques (see Appendix 
\ref{thebulkder}) that
\begin{eqnarray}
W_{d}^{\sigma}(r,b|\Lambda) &=& W_{d}^{\sigma}(0,b|\Lambda)+ 
\frac{\pi}{(4\pi)^{d/2}\Gamma(d/2)\sin(\pi d/2)} r^{d/2-1}\nonumber 
\\
& &+ b \frac{\pi(d+\sigma-2)}{2(4\pi)^{d/2}\Gamma(d/2)\sin(\pi( 
d+\sigma)/2)} r^{d/2-1+(\sigma-2)/2}\nonumber \\
& & +O(r^{(d+2(\sigma-2))/2-1},r ),
\end{eqnarray}
where $2<\sigma<4$, $2<d<4$ and $d+\sigma<6$.

Since $d+\sigma=6$ includes the especially important case $d=\sigma=3$ 
we also present the corresponding result for that case
\begin{eqnarray}
W_{d}^{\sigma}(r,b|\Lambda) &=& W_{d}^{\sigma}(0,b|\Lambda)+ 
\frac{\pi}{(4\pi)^{d/2}\Gamma(d/2)\sin(\pi d/2)} r^{d/2-1}\nonumber \\
& &- b \frac{2}{(4\pi)^{d/2}\Gamma(d/2)} r\ln{r}+O(r).
\end{eqnarray}

Inserting these expansions into the equation of state (\ref{es}) we 
obtain
\begin{equation}
x_{1} = X^{sr}(x) + b L^{2-\sigma} X^{lr}(x) + \frac{x_{2}^{2}}{x^{2}}
+ \Lambda \! - \! \mbox{dependent terms} +O \left(xL^{-(4-d)}, x^{d/2 + \sigma 
-3}L^{-2(\sigma -2)} \right),
        \label{esscaling}
\end{equation}
where $x=rL^2$, $x_1=(K-K_c)L^{1/\nu}$, $x_2=hL^{\Delta/\nu}/\sqrt{K}$ 
with $\nu=1/(d-2)$, $\Delta=(d+2)/(2(d-2))$ (see \cite{BDT00}) and
\begin{equation}
X^{sr}(x)=\frac{1}{x}+I_{{\rm 
scaling}}^0(x,d)+\frac{\pi}{(4\pi)^{d/2}\Gamma(d/2)\sin(\pi 
d/2)} x^{d/2-1},
\label{ssrterm}
\end{equation}
\begin{eqnarray}
X^{lr}(x) &=& 
\frac{\pi(d+\sigma-2)}{2(4\pi)^{d/2}\Gamma(d/2)\sin(\pi( 
d+\sigma)/2)} x^{d/2-1+(\sigma-2)/2}\nonumber \\
& & -(1+x\frac{\partial}{\partial x}) I_{{\rm 
scaling}}^{(\sigma-2)/2}(x,d).
\label{slrterm}
\end{eqnarray}

To analyze the equation of state and the behavior of quantities such 
as the (reduced) magnetisation $m=h/r$, the susceptibility $\chi=1/r$ 
\cite{BDT00} the information required, in addition to that given 
above, is with respect to the asymptotics of $I_{{\rm 
scaling}}^0(x,d)$ and $I_{{\rm scaling}}^{(\sigma-2)/2}(x,d)$ in 
different regions of the thermodynamic parameters.  We will be 
interested in the behavior of $m$ and $\chi$ slightly above, in the 
region of, and below the critical temperature.

The asymptotics of $I_{{\rm scaling}}^0(x,d)$ are well known
\begin{equation}
I_{{\rm scaling}}^0(x,d) \simeq-\frac{1}{x} 
+\frac{d\sqrt{2}}{\pi^{(d-1)/2}}x^{(d-3)/4}e^{-\sqrt{x}}, \ \ \ \ x 
\rightarrow \infty
\label{xgoestoinfinity}
\end{equation}
and
\begin{equation}
I_{{\rm scaling}}^0(x,d) \simeq I_{\rm scaling}^{0}(0,d), \ \ \ \ \ x 
\rightarrow 0
\label{xgoesto0}
\end{equation}
where
\begin{eqnarray}
I_{\rm scaling}^{0}(0,d) &=& \int_{0}^{\infty} dt \left[\sum_{\vec{k} 
\neq 0}e^{-4 \pi^{2}k^{2} t} - \left(4 \pi t \right)^{-d/2} \right] 
\nonumber \\
&=& \int_{0}^{\infty} dt \left[
 \left(4 \pi t \right) ^{-d/2} \sum_{\vec{k} \neq 0} 
e^{-k^{2}/4 t} -1\right]  \nonumber \\
& \equiv & D_{0}
        \label{express1}
\end{eqnarray}

The corresponding asymptotics of $I_{{\rm 
scaling}}^{(\sigma-2)/2}(x,d)$ are (see Appendix \ref{lras})
\begin{equation}
I_{{\rm scaling}}^{p}(x,d) \simeq C_p x^{-2}, \ \ x \rightarrow 
\infty,
\end{equation}
where
\begin{equation}
C_{p} = -\frac{(1+p) 
4^{1+p}}{\pi^{d/2}}\frac{\Gamma(1+p+d/2)}{\Gamma(-p)}\sum_{\vec{k} 
\neq 0}\frac{1}{k^{d+2(p+1)}}
        \label{Cpdef}
\end{equation}
and
\begin{eqnarray}
I_{{\rm scaling}}^{p}(x,d)\simeq I_{{\rm scaling}}^{p}(0,d)& = & 
\frac{1}{\Gamma(1-p)}\int_0^\infty t^{-p} \left[ (4\pi 
t)^{-d/2}\sum_{{\bf k}\ne {\bf 0}}e^{-k^2/4t}-1\right] dt \\
&=& \frac{1}{\Gamma(1-p)}\int_0^\infty t^{-p} \left[ 
\sum_{{\bf k}\ne {\bf 0}}e^{-4\pi^2k^2 t}-(4\pi 
t)^{-d/2}\right] dt.  \nonumber
\end{eqnarray}
Obviously the right hand sides are well defined for $p<1$ and $2<d<4$ 
both around the lower and the upper limit of integration.  We will 
denote this constant by $D_p$.

\section{Finite size effects and the susceptibility}
\label{sec:chi}

Given the equation of state, we are now in a position to explore the 
behavior of various thermodynamic properties of the system with 
sub-leading long range interactions. Here, we look at the 
susceptibility of such a system. We first consider the case 
$2< d <4$ and $2 < \sigma < 4$. Furthermore, we assume $d + \sigma < 6$. 
The scaling form of the equation of state is
\begin{equation}
x_{1} = X^{sr}(x) + b L^{2-\sigma} X^{lr}(x) + 
\frac{x_{2}^{2}}{x^{2}} + \Lambda \!  
- \!  \mbox{dependent terms} +O \left(xL^{-(4-d)}, x^{d/2 + \sigma 
-3}L^{-2(\sigma -2)} \right)
        \label{eosb}
\end{equation}
Here, $x=rL^{2}$, and the susceptibility $\chi$ is given by
\begin{equation}
\chi = \frac{1}{r}
        \label{chidef}
\end{equation}
assuming that $h=0$. We now analyze the behavior of $\chi$ for 
temperature, $T$, close the critical temperature, above and below, and 
for $T$ in the immediate vicinity of the critical temperature, in that 
finite size rounding is evident.

To that end, we need the asymptotics of $X^{sr}(x)$ and $X^{lr}(x)$ in 
the limits of large and small $x$. Making use of the asymptotics of 
$I^{0}_{\rm scaling}(x,d)$ and $(1+ x \partial/\partial x)I^{p}_{\rm 
scaling}(x,d)$, we have
\begin{equation}
X^{sr}(x) \simeq \left\{ \begin{array}{ll} 
\frac{d\sqrt{2}}{\pi^{(d-1)/2}}x^{(d-3)/4}e^{-\sqrt{x}} 
+A_{d}x^{d/2-1} , & x \rightarrow \infty \\
\frac{1}{x} +D_{0} , & x \rightarrow 0 \end{array} \right.
        \label{Xsr}
\end{equation}
and
\begin{equation}
X^{lr}(x) \simeq \left\{ \begin{array}{ll} B_{d, \sigma}x^{(d+ 
\sigma)/2 -2} + C_{(\sigma-2)/2}x^{-2} , & x \rightarrow \infty \\
D_{(\sigma-2)/2} , & x \rightarrow 0 \end{array} \right.
        \label{Xlr}
\end{equation}
where
\begin{equation}
A_{d}= \frac{\pi}{\left(4 \pi \right)^{d/2} \Gamma(d/2) \sin \left( 
\pi d /2 \right)} <0
        \label{Ad}
\end{equation}
and
\begin{equation}
B_{d, \sigma} = \frac{\pi \left( d +\sigma - 2 \right)}{2 \left( 4 
\pi \right)^{d/2} \Gamma(d/2) \sin \left( \pi (d+2)/2 \right)} > 0
        \label{Bdsig}
\end{equation}

We begin with the case $T=T_{c}$. The Eq. (\ref{eosb}) becomes
\begin{equation}
0 = X^{sr}(x) + b L^{2- \sigma} X^{lr}(x)+\cdots.
        \label{eosc}
\end{equation}
Let $x_{0}$ be the solution of the equation $X^{sr}(x_{0}) =0$. 
Obviously, $x_{0}$ which is $O(1)$, is a positive constant. Taking 
into account that $\sigma >2$ and solving Eq. (\ref{eosc}) 
iteratively, we obtain
\begin{equation}
\frac{1}{x} \simeq \frac{1}{x_{0}} + b L^{2 - \sigma} 
X^{lr}(x_{0})/\left(x_0^2 X^{\prime sr}(x_{0})\right)
        \label{eosd}
\end{equation}
where $X^{\prime sr}(x_{0})$ is the derivative of  $X^{sr}(x)$ at 
$x=x_{0}$. Recalling that $\chi =1/r$ and $x=rL^{2}$, one immediately 
obtains from (\ref{eosd})
\begin{equation}
\chi \simeq x_{0}^{-1}L^{2} + bL^{2-(\sigma -2)}X^{lr}(x_{0})/
\left(x_0^2X^{\prime 
sr}(x_{0})\right)
        \label{eose}
\end{equation}
It is clear that in a Monte Carlo simulation if neither $\sigma$ nor 
$L$ is particularly large, then the correction terms in (\ref{eose}), 
which go as $L^{2-(\sigma -2)}$ might well be as large, numerically, 
as the leading order terms which scale as $L^{2}$, depending, of 
course, on the values of $b$, $X^{lr}(x_{0})$ and $X^{ \prime 
sr}(x_{0})$. 

Let us now consider the case in which $T$ is fixed close to, but also 
\emph{above} $T_{c}$ and $L \rightarrow \infty$. Then, $x_{1} 
\rightarrow - \infty$, and, taking into account the corresponding 
asymptotic behavior of $X^{sr}$ and $X^{lr}$ for $x \rightarrow 
\infty$ ($rL^{2} \gg 1$) we can rewrite Eq. (\ref{eosb}) in the 
following form
\begin{equation}
K-K_{c}\simeq A_{d}r^{d/2 -1}+bB_{d,\sigma}r^{d/2-1 + \sigma/2 
-1}+bC_{p}L^{-(d + \sigma)}r^{-2}
        \label{eosf}
\end{equation}
Solving this equation iteratively, we obtain
\begin{equation}
\chi \simeq \chi_{0} \left\{ 1 + b \gamma \chi_{0}^{-(\sigma -2)/2} \left[ 
\frac{B_{d, \sigma}}{A_{d}} + C_{(\sigma -2)/2} \left( \chi_{0} 
L^{-2} \right)^{(d+\sigma)/2} \right] \right\}
        \label{chisol1}
\end{equation}
where $\chi_{0}$ is the susceptibility of the corresponding infinite 
system with short-range interactions only, i.e.
\begin{equation}
\chi_{0} = \left( \frac{A_{d}}{K-K_{c}}\right)^{\gamma}
        \label{chisol2}
\end{equation}
with 
\begin{equation}
\gamma = \frac{2}{d-2}
        \label{gammasrdef}
\end{equation}
The above solution is valid when $rL^{2} \gg 1$, i.e. $L^{2} \gg 
\chi_{0}$. Note that the dominant finite-size corrections to the 
behavior of the total susceptibility are of order $L^{-(d+\sigma)}$. 
That is, they are not exponentially small, nor are they 
cutoff-dependent. The existence of corrections of this type in the case 
of leading-order long-range interactions is well known. First they have 
been derived in the framework of the spherical model \cite{SP89,BD91}.
Analogous is also the behavior of the $O(n)$ model within $\epsilon =2\sigma
-d$ expansion \cite{CT2000} (at least up to the first order in $\epsilon $). 

Finally, let us consider the case $T < T_{c}$. Then, $x_{1} \rightarrow 
\infty$, which leads to $x \rightarrow 0$. Eq. (\ref{eosb}) becomes
\begin{equation}
K-K_{c} \simeq \frac{1}{rL^{d}} +D_{0}L^{2- d} + bL^{4- (d + 
\sigma)}D_{(\sigma -2)/2} + \frac{h^{2}}{Kr^{2}} .
        \label{eosg}
\end{equation}
In the absence of an external field, the iterative solution of the 
above equation yields
\begin{equation}
\chi \simeq \left(K-K_{c}\right) L^{d}- D_{0}L^{2} -bD_{(\sigma - 
2)/2}L^{4- \sigma}
        \label{chisol3}
\end{equation}

Now, we turn to the case $d + \sigma =6$ ($2 < d < 4$, $2 < \sigma < 
4$). This is especially \emph{a propos}, in light of the fact that the 
van der Waals interaction in three dimensions leads to a contribution 
in which $d= \sigma = 3$. In this case, it is necessary to take into 
account the special form of $X^{lr}$:
\begin{equation}
X^{lr}(x) =2B x \ln L -B x \ln x + \left(1+x\frac{\partial}{\partial x}\right) 
I^{(4-d)/2}_{\rm scaling}(x,d) ,
        \label{Xlrspec}
\end{equation}
where
\begin{equation}
B = \frac{2}{\left( 4 \pi \right)^{d/2} \Gamma(d/2)}.
        \label{Bdef}
\end{equation}
The first term is responsible for the leading-order finite-size
corrections that are due to the subleading long-range part of the interaction.

Proceeding as in the case $d+ \sigma <6$, it is readily demonstrated 
that \begin{enumerate}
\item[a)] For $T=T_{c}$:
\begin{equation}
\chi \simeq x_{0}^{-1}L^{2} + \frac{2 b B}{x_0 X^{\prime 
sr}(x_{0})}L^{d-2} \ln L .
        \label{newchi}
\end{equation}
The correction term is obviously important in the analysis of Monte 
Carlo data.
\item[b)] For $T >T_{c}$:
\begin{equation}
\chi \simeq \chi_{0} \left[ 1 - b \chi_{0}^{-(4-d)/2} 
\left(\frac{B}{\left(\frac{d}{2} -1\right) A_{d}} \ln 
\frac{1}{\chi_{0}} - C_{p} \left( \chi_{0} L^{-2} \right)^{3} 
\right) \right] .
        \label{newchi1}
\end{equation}

\item[c)] For $T < T_{c}$:

In this region, the corrections to bulk behavior due to long-range 
corrections play no role, and the solution remains unaltered.

\end{enumerate}

\section{Conclusions}
\label{sec:dis}

In this paper, we have reported the results of an investigation into 
the critical point properties of a finite spherical model in which 
interactions contain a component that is long-range, but 
insufficiently so to alter the asymptotic singularities of its 
thermodynamics---in particular the critical exponents.  One can 
envision interactions decaying as $r^{-(d+\sigma)}$, where $d$ is the 
dimensionality of the system and $\sigma>2$.  An important example is 
van der Waals interaction, which decays in three dimensions as 
$r^{-6}$ that is consistent with $\sigma=3$.  The finite system that 
we consider is subject to periodic boundary conditions, and, thus, 
provides a model for the sorts of systems that are studied in computer 
simulations.  This investigation was stimulated by recent work of Chen 
and Dohm, \cite{C&D1,C&D2,C&D3} in which a combination of a spin-spin 
interaction truncated in momentum-space and a sharp momentum-space 
cutoff on fluctuations gives rise to an effectively long-range 
interaction.

In the critical region we find that the susceptibility of the finite
system $\chi(t,h;L)$ is of the form (see Eq. (\ref{eosb}),
(\ref{chidef}),
(\ref{eose}))
\begin{equation}
 \chi(t,h;L)=L^{\gamma/\nu}Y(x_1,x_2,bL^{2-\sigma}), 
\end{equation}
or, equivalently,
\begin{equation}
 \chi(t,h;L)=L^{\gamma/\nu}
\left[Y^{sr}(x_1,x_2) + bL^{2-\sigma} Y^{lr}(x_1,x_2)\right],
\end{equation}
where $x_1=c_1tL^{1/\nu}$, $x_2=c_2hL^{\Delta/\nu}$, and $Y$, $Y^{sr}$ 
and $Y^{lr}$ are universal functions.  The quantities $c_1$, $c_2$ and 
$b$ are nonuniversal constants.  Note, that the above structure of the 
finite-size scaling function in systems with subleading long-range 
interactions is different from the corresponding one for systems with 
essentially finite range of interaction \cite{F72,FB72,B83,P90,BDT00}.  
The new length scale which is involved does not lead to corrections of 
the finite-size scaling picture known before, but leads, see below, to 
leading finite-size contributions above $T_c$.

In the range of parameters, for which $tL^{1/\nu}=O(1)$, and also a
bit below the critical point, where $tL^{1/\nu}\rightarrow -\infty$,
the long-range contributions represented by $Y^{lr}$ are merely corrections to
the leading finite-size behavior. Somewhat more interestingly---and of 
greater practical significance---there are also corrections to the 
dependence on system size of singular thermodynamic properties at the 
critical point that can conceivably cloud the numerical analysis of 
Monte Carlo data, in that the corrections, while less important in an 
asymptotically large system, may be of the same order of magnitude in 
systems that are a realizable size (see Eqs. (\ref{eose}),
(\ref{chisol3}) and (\ref{newchi})).  The studies reported here ought 
to provide, at the very least, a conceptual basis for the critical 
evaluation of Monte Carlo results.

On the other hand, in the high-temperature, unordered phase, where 
$tL^{1/\nu}\rightarrow\infty$, we find that the long-range portion of 
the interaction between spin degrees of freedom gives rise to 
contributions of the order of $bL^{-(d+\sigma)}$ that swamp the 
exponentially small terms that are expected to characterize the 
signature of finite size in systems with periodic boundary conditions 
and short range interactions.  In other words the {\it subleading} 
long-range part of the interaction gives rise to a {\it dominant} 
finite-size dependence in this regime.  This is entirely consistent 
with the inherent long-range correlations that attend long-range 
interactions, but it goes beyond the standard finite-size scaling 
formulation.  More explicitly, one obtains $Y^{sr}(x_1,0)\sim 
\exp(-{\rm const.}  \ x_1^\nu)$, while
\begin{equation}
  Y^{lr}(x_1,0)\sim x_1^{-(d+2)\nu-\gamma},
\label{aY}
\end{equation}
when $x_1\rightarrow\infty$.  This asymptotic follows from the 
requirement the finite-size corrections to be of the order of 
$L^{-(d+\sigma)}$ in this regime, which is to be expected on general 
grounds and is supported from the existing both exact and perturbative 
results for models with {\it leading} long-range interaction included 
\cite{SP89,BD91,CT2000}.  Note that (\ref{aY}) implies for the 
temperature dependence of this corrections that
\begin{equation}
 \chi(t,h;L)-\chi(t,h;\infty)\sim t^{-(d+2)\nu-\gamma}L^{-(d+\sigma)},
\ tL^{1/\nu}\rightarrow\infty. 
\end{equation}
This prediction is in full agreement with Eq.  (\ref{chisol1}).  
Obviously, the existence of such power-law finite-size dependent 
dominant terms above $T_c$ is of crucial significance in the analysis 
of the Monte Carlo data from simulation of such systems.

Finally, it is worth noting that the system considered here is 
equivalent to an $O(n)$ dimensional vector spin model in the limit 
$n\rightarrow \infty$ \cite{BDT00}.  Because of the spin-wave 
excitations, the bulk correlation length of such a system is 
identically infinite below $T_c$ (for any $O(n)$, $n \neq 1$ model).  
As a result the correlations decay in a power-law in this regime.  The 
direct spin-spin interaction decays faster there, and that is why for 
$T<T_c$ we obtain no essential finite-size contributions due to the 
subleading term in the Fourier transform of the interaction.  The 
situation is different in Ising-like systems.  There below $T_c$ the 
correlation length is finite, the correlations decay exponentially 
fast in a system with only a $q^2$ term in the Fourier transform of 
the interaction.  Since, when the interaction is long-ranged the 
correlations cannot decay faster than the corresponding direct 
spin-spin interaction, one should expect modifications of scaling of 
the type presented in the current article for $T>T_c$ to be necessary 
for Ising-like systems also {\it below} $T_c$.

In addition, in the current article we have proven the 
approximation formula (see Eq. (\ref{cd14}))
\begin{eqnarray}
\sum_{n=-M}^{M-1}e^{-4\pi ^2tn^2} &\simeq&\frac 1{\sqrt{4\pi t}}\left[ 
\mathop{\rm erf}
\left( \sqrt{4\pi ^2t}M\right) +\sum_{k\neq 0}e^{-k^2/4t}\right
]  \nonumber
\\
&&-\frac{4 \pi^{2}}{3}Mte^{-4\pi ^2tM^2}
\end{eqnarray}
which is of a bit more general mathematical interest and which also 
might be useful in a lot of studies of finite-size effects by exact or 
perturbative methods.

\section*{Acknowledgements}

D. D. thanks Drs.  N. S. Tonchev and J. G. Brankov for a critical 
reading of the manuscript and acknowledges the hospitality of UCLA 
while the work reported here was performed .  J. R. acknowledges the 
support of NASA through grant number NAG3-1862.

\appendix
\section{Derivation of the central result of Chen and Dohm}
\label{app:derivation}

The expression in which the size dependence of the statistical 
mechanics appears in the papers of Chen and Dohm \cite{C&D1,C&D2,C&D3} 
is given by the difference between a sum over the set of allowed wave 
vectors in a hypercubic system having a linear extent $L$ in every 
direction and the integral for such a system in the limit $L=\infty$.  
It is assumed that both the sum and the integral are taken over a 
region of wave-vector space that is also a $d$-dimensional cube.  The 
system under consideration is subject to periodic boundary conditions 
in all dimensions.  This paper addresses the question of the source of 
the violation of scaling  found  by Chen and Dohm by focusing on 
the effective long-range nature of the interactions that are generated 
by the combination features assumed to hold for the system considered 
by them.  However, for the reader interested in looking at their 
papers on the subject we provide here details of the derivation of the 
terms in the equation of state that fall off as a power in the size, 
$L$, of the system.

The quantity from which the power-law finite-size corrections arise 
is the difference between a lattice sum over wave-vectors, $\vec{k}$ 
and the integral to which it is equal in the thermodynamic limit. This 
difference, which is introduced, for instance, in Eq. (5) of 
\cite{C&D3}, is given by
\begin{equation}
\tilde{\Delta}_1\left( \Lambda ,\chi ^{-1}\right) =I\left( \Lambda ,\chi
^{-1}\right) -S_L\left( \Lambda ,\chi ^{-1}\right),
\label{cd1}
\end{equation}
where 
\begin{equation}
I\left( \Lambda ,\chi ^{-1}\right) =\frac 1{\left( 2\pi \right) ^d}
\int_{-\Lambda }^\Lambda dk_1\cdots \int_{-\Lambda }^\Lambda dk_d\frac 1{
\chi ^{-1}+\vec{k}^2}
\label{cd2}
\end{equation}
and
\begin{equation}
S_L\left( \Lambda ,\chi ^{-1}\right) =L^{-d}\sum_{\vec{n}\neq 0}\frac 1{\chi
^{-1}+\left( \frac{2\pi }L\right) ^2\vec{n}^2}
\label{cd3}
\end{equation}
with
\begin{equation}
n_i\in [-M,M),i=1,\cdots d,M=L\Lambda /(2\pi ).
\label{cd4}
\end{equation}

We start by looking at the term $I(\Lambda, \chi^{-1})$. This term 
yields readily to analysis. Making use of the identity
\begin{equation}
\int_0^\infty \exp \left( -ax\right) =\frac 1a, \mathop{\rm Re} a>0,
\label{cd5}
\end{equation}
and
\begin{equation}
\frac 1{2\pi }\int_{-\Lambda }^\Lambda \exp \left( -ak^2\right) dk=\frac{%
\mathop{\rm erf} \left[ \Lambda \sqrt{t}\right] }{\sqrt{4\pi t}},
\label{cd6}
\end{equation}
the function $I\left( \Lambda ,\chi ^{-1}\right) $ can be rewritten in the form
\begin{equation}
I\left( \Lambda ,\chi ^{-1}\right) =\int_0^\infty \left( 4\pi t\right)
^{-d/2}\exp \left[ -t\chi ^{-1}\right]
\left\{ 
\mathop{\rm erf}
\left[ \Lambda \sqrt{t}\right] \right\} ^d dt.
\label{cd7}
\end{equation}

We now turn to the expression for the sum, $S_{L}( \Lambda, 
\chi^{-1})$. This sum is given by
\begin{eqnarray}
S_L\left( \Lambda ,\chi ^{-1}
\right) &=&L^{-\left( d-2\right) }\sum_{\vec{n}
\neq 0}\frac 1{\chi ^{-1}L^2+4\pi ^2\vec{n}^2}  \nonumber \\
&=&L^{-\left( d-2\right) }\int_0^\infty e^{-\chi ^{-1}L^2t}\left(
\sum_{n=-M}^{M-1}e^{-4\pi ^2tn^2}\right) ^d dt -L^{-d}\frac 
{1}{\chi^{-1}}.  \label{cd8}
\end{eqnarray}
We now write
\begin{equation}
S\left( t,M\right) =\sum_{n=-M}^{M-1}e^{-4\pi ^2tn^2}.
\label{cd9}
\end{equation}
and apply the Poisson summation formula
\begin{equation}
\sum_{n=a}^bf\left( n\right) =\sum_{k=-\infty }^\infty \int_a^be^{i2\pi
kn}f\left( n\right) dn+\frac 12\left[ f\left( a\right) +f\left( b\right)
\right]
\label{cd10}
\end{equation}
This yields
\begin{eqnarray}
S(t,M) &=& \sum_{k=-\infty}^{\infty} \int_{-M}^{M}e^{-4 \pi tn^{2} +2 
\pi i k n} \ dn \nonumber \\
&=& \frac{1}{\sqrt{4 \pi t}} \mathop {\rm erf} \left( \sqrt{4 
\pi^{2}t} M \right) + 2 \sum_{k=1}^{\infty}\int_{-M}^{M} e^{- 4 
\pi^{2}n^{2}t} \cos 2 \pi k n \ d n
        \label{cd11}
\end{eqnarray} 
If the only term retained in our result for $S(t,M)$ were the first 
one on the last line of (\ref{cd11}), then there would be perfect 
cancellation between the sum and the integral. The difference between 
the two results from the second term. We now construct an asymptotic 
expansion for that term. We have
\begin{eqnarray}
2 \sum_{k=1}^{\infty}\int_{-M}^{M} e^{- 4 \pi^{2}n^{2}t} \cos 2 \pi k 
n \ dn &=& 2 \sum_{k=1}^{\infty} \left\{\int_{- \infty}^{\infty}e^{- 4 
\pi^{2}n^{2}t} \cos 2 \pi k n \ dn - 2 \int_{M}^{\infty} e^{- 4 
\pi^{2}n^{2}t} \cos 2 \pi k n \ dn  \right\}\nonumber \\
&=& 2\sum_{k=-\infty}^{\infty} \left\{ \frac{1}{\sqrt{4 \pi t}} e^{- 
k^{2}/4t} - 2 \Re \left[ \int_{M}^{\infty} e^{-4 \pi^{2}t n^{2} +2 
\pi i k n} dn \right] \right\} \nonumber \\
& = & 2\sum_{k=-\infty}^{\infty} \left\{ \frac{1}{\sqrt{4 \pi t}} e^{- 
k^{2}/4t} - 2 \Re \left[ \int_{0}^{\infty} e^{-4 \pi^{2}t (M 
+n^{\prime})^{2} +2 \pi i k (M+n^{\prime})} dn^{\prime} \right] 
\right\} \nonumber \\
&\approx & 2\sum_{k=-\infty}^{\infty} \left\{ \frac{1}{\sqrt{4 \pi t}} e^{- 
k^{2}/4t} - 8 M t e^{- 4 \pi^{2} t M^{2}} 
\frac{1}{k^{2} + 16 \pi^{2} t^{2} M^{2}} \right\} \nonumber \\
\label{cd12}
\end{eqnarray}
We obtain the last line in (\ref{cd12}) by retaining only those terms 
in the exponent in the next-to-last line that are linear in the 
integration variable $n^{\prime}$. 

It is now fairly straightforward to demonstrate that the combination 
$k^{2} + 16 \pi^{2} t^{2} M^{2}$ in the last line of (\ref{cd12}) can 
be replaced by $k^{2}$ with no loss of accuracy in the evaluation of 
the sum.  Making use of the result
\begin{equation}
\sum_{k=1}^{\infty}\frac{1}{k^{2}} = \frac{\pi^{2}}{6}
        \label{cd13}
\end{equation}
we end up with
\begin{eqnarray}
\sum_{n=-M}^{M-1}e^{-4\pi ^2tn^2} &=&\frac 1{\sqrt{4\pi t}}\left[ 
\mathop{\rm erf}
\left( \sqrt{4\pi ^2t}M\right) +\sum_{k\neq 0}e^{-k^2/4t}\right
]  \nonumber
\\
&&-\frac{4 \pi^{2}}{3}Mte^{-4\pi ^2tM^2}
\label{cd14}
\end{eqnarray}

The remainder of the calculation involves the insertion of the above 
results into the expression (\ref{cd8}). The key term arises from a 
cross-term in the expansion of of the $d^{\rm th}$ power of 
(\ref{cd14}). That term is 
\begin{equation}
- \frac{d}{3} 4 \pi^{2}t M \left( 4 \pi t \right)^{-(d-1)/2} \left[ 
\mathop {\rm erf} \left( \sqrt{4 \pi^{2}t}M \right) \right]^{(d-1)} 
e^{-4 \pi^{2} t M^{2}}
        \label{cd15}
\end{equation}
The remainder of the calculation involve scaling the system size, $L$ 
out of the integral over $t$ in (\ref{cd8}). To recover the form 
exhibited in \cite{C&D2}, on makes use of the equality
\begin{equation}
(4 \pi t)^{-(d-1)/2} \left[ \mathop {\rm erf} \left( \sqrt{4 
\pi^{2}t}M \right) \right]^{d-1} = \left[ \frac{1}{2 \pi} 
\int_{-M}^{M}e^{-tk^{2}}dk \right]^{d-1}
        \label{cd16}
\end{equation}
and of the Jacobi identity
\begin{equation}
\sum_{\vec{k}\neq 0} e^{-k^{2}/4t} = \left[ \sqrt{4 \pi 
t}\sum_{k}e^{-4 \pi k^{2}t}\right]^{d}-1
        \label{cd17}
\end{equation}
which leads to the end-result
\begin{equation}
S_{L}\left( \Lambda, \chi^{-1}\right) = I\left(\Lambda, 
\chi^{-1}\right)-L^{-(d-2)}I_{1}\left( L^{2} \chi^{-1}\right) - 
\Lambda^{d-2}a_{1}\left(d, \chi^{-1}\Lambda^{-2}\right) \left( 
\Lambda L \right)^{-2} + \cdots
        \label{cd18}
\end{equation}
which leads to 
\begin{equation}
\tilde{\Delta}_{1} \left( \Lambda, \chi^{-1} \right) = 
L^{-(d-2)}I_{1}\left( L^{2}\chi^{-1}\right) + \Lambda^{d-2} 
a_{1}\left( d, \chi^{-1}\Lambda^{-2}\right) \left( \Lambda L 
\right)^{-2}
        \label{cd19}
\end{equation}
with the functions $I_1$ and $a_{1}$ as given by Eqs. 
(7) and (8) in \cite{C&D3}, 
respectively. Note that $I_1(x)=I^0_{{\rm scaling}}(x,d)$, and
$a_1(d,x)=-I^{0}_{{\rm cut \ off}}(x)/(4\pi^2)$.

\section{The combined influence of an interaction going as ${ q}^{2}$ 
and a sharp cutoff in $q$-space}
\label{app:cutoff}

Consider an interaction that Fourier transforms to $q^{2}$ exactly.  
That is, imagine that the Fourier transform on the interaction is as 
given by Eq.  (\ref{expand1}) with $R(\vec{q})=0 $.  The first term in 
this expansion gives rise to a real-space interaction that is entirely 
local.  We thus focus on the term that goes as $q^{2}$.  For 
simplicity, we start by restricting our attention to a one-dimensional 
system.  If the interacting spins reside on a lattice with unit 
lattice spacing the form of this interaction in real space is given by
\begin{eqnarray}
\int_{- \pi}^{\pi}q^{2}e^{iqn}dq &=& -\frac{d^{2}}{dn^{2}} \int_{- 
\pi}^{\pi}e^{iqn}dq \nonumber \\
&=& -\frac{d^{2}}{dn^{2}} 2 \frac{\sin n \pi}{n} \nonumber \\
&=& \left\{ \begin{array}{ll} \frac{2}{3} \pi^{3} & n=0 \\ 4 
\frac{\cos \pi n}{n^{2}} & n \neq 0 \end{array} \right.
        \label{resp1}
\end{eqnarray}
As $n$ is an integer, the real space interaction decays as a modulated 
power law.  Such an interaction has a different range than one in 
which the power law is ``pure,'' in that there is no alternation in 
the sign of the interaction as the distance between the spins 
increases.  However, this system displays long-range correlations, 
which manifest themselves in the size-dependence of thermodynamic 
quantities.

One can assess the impact of this interaction on, say, the free energy 
by separating it into a short-range piece and a long range one.  At 
one extreme, one can take the short range piece to be the $n=0$ part 
of the interaction.  If one retained that term and discarded the 
long-range portion of the interaction, one would be left with a system 
in which there are no correlations between degrees of freedom at any 
non-zero separation.  Suppose we start with this approximation, which 
is not too far from reality at very high temperatures.  Then, we 
discover how the long-range portion of the interaction influences a 
system with periodic boundary conditions with the use of the 
perturbation expansion.  Writing the Hamiltonian of this system in 
the form
\begin{equation}
{\cal H} = H_{0} + H_{\rm lr}
        \label{hdiv1}
\end{equation}
where $H_{0}$ is the short-range portion of the interaction, $\propto 
\sum_{i}s_{i}^{2}$, and $H_{\rm lr}$ is the long-range portion, we 
write the free energy as follows
\begin{eqnarray}
{\cal F} &=& -\ln \left[ \sum_{\left \{ s_{i} \right \}} e^{-{\cal H}} 
\right] \nonumber \\
&=& - \ln \left[ \sum_{\left \{ s_{i} \right \}}e^{ -H_{0} -H_{ \rm 
lr}}\right] \nonumber \\
&=& - \ln \left[ \sum_{\left\{ s_{i} \right \} }e^{-H_{0}}\right] + 
\frac{\sum_{ \left\{ s_{i} \right\}} e^{-H_{0}} H_{\rm lr}}{\sum_{ 
\left\{ s_{i} \right\} }e^{ -H_{0}}} + \cdots \nonumber \\
& = & F_{0} + \langle H_{\rm lr} \rangle + \cdots
\label{Hexp1}
\end{eqnarray}
The first term on the last line of (\ref{Hexp1}) is the free energy of 
the system with short-range interactions only.  The second term is the 
result of expanding the free energy to first order in the long-range 
portion of the interaction.  This second term takes the form
\begin{equation}
\sum_{n}V(n)\langle s_{m}s_{m+n}\rangle
        \label{firstorder1}
\end{equation}
where, $V(n)$ is the real space version of the long-range interaction, 
as given in Eq.  (\ref{resp1}).

Let's imagine the case of a system with periodic boundary conditions.  
Such a system is equally well represented by an infinite set of 
duplicates of a the finite system.  In this case, the correlation 
function $\langle s_{m} s_{m+n} \rangle$ is equal to zero unless $n=0$ 
or $n= \pm k L$, where $k$ is an integer and $L$ is the size of the 
system.  We take $L$ to be an even integer.  One then obtains for the 
influence of the long range interaction on the free energy of the 
system
\begin{equation}
8\langle s_{m}^{2}\rangle \sum_{k=1}^{\infty}\frac{1}{k^{2}L^{2}}
        \label{firstorder2}
\end{equation}
Note that this influence goes as a power law in the size of the 
periodically continued one-dimensional system.  It is interesting to 
note that the detailed dependence on the size of the system is 
different from the above when $L$, the system's size in terms of the 
distance between sites, is an odd integer.

In three dimensions, the one-dimensional Brilouin zone is replaced by 
a cubic zone in reciprocal space, and the interaction  
in real space is given by
\begin{equation}
\int_{-\pi}^{\pi} dq_{x}\int_{-\pi}^{\pi} dq_{y}\int_{-\pi}^{\pi} 
dq_{z} \left( q_{x}^{2} +q_{y}^{2} +q_{z}^{2} \right) e^{i\vec{q} 
\cdot \vec{n}}
        \label{resp2}
\end{equation}
It is straightforward to see that the interaction consists of three 
contributions, each long range in one direction and extremely short 
range in the two others.  For example
\begin{equation}
\int_{-\pi}^{\pi} dq_{x}\int_{-\pi}^{\pi} dq_{y}\int_{-\pi}^{\pi} 
dq_{z}q_{x}^{2}e^{-\vec{q} \cdot \vec{n}} = V(n_{x}) \delta_{n_{y}} 
\delta_{n_{z}}
\label{resp3}
\end{equation}
The overall interaction is thus long range, but highly anisotropic.  
That is, a given spin interacts with spins arbitrarily far away, but 
only with spins separated from it by a displacement vector that points 
entirely along the $x$, $y$ or $z$ axis.

\section{The effect of truncation of the Fourier transform of the 
interaction}
\label{app:truncation}

In Appendix \ref{app:cutoff}, it is established that an interaction 
whose Fourier transform is truncated at the quadratic term, coupled 
with a sharp cutoff in Fourier space, has, in real space, a modulated 
power-law tail.  In this and the following appendix, we demonstrate 
that both the truncation and a sharp cutoff are required for that 
long-range behavior to be manifested.  Here, we investigate the effect 
of truncation only.  As in the previous appendix, we focus on one 
dimension.  There is every reason to believe that the our conclusions 
are unaltered in higher dimensionality.

We will limit our discussion to the sum over wave vectors entering 
into the spherical model equation of state.  First, we evaluate that 
sum for a lattice system with nearest neighbor interactions.  In that 
case, the Fourier transform is of the form $\cos q$, where $q$ is the 
wave vector.  We assume unit lattice spacing.  The sum of interest has 
the form
\begin{equation}
\sum_{i=1}^{N}\frac{1}{\alpha - \cos q_{i}}
        \label{sum1b}
\end{equation}
where
\begin{equation}
q_{i} = \frac{2 \pi i}{N}
        \label{ki}
\end{equation}
This sum can be recast as a contour integral.  Write
\begin{equation}
\cos q = \frac{1}{2} \left( z + \frac{1}{z}\right)
        \label{cos1}
\end{equation}
where $z$ lies on the unit circle.  The $z$'s appearing the sum in 
(\ref{sum1b}) are the $N$ values of the $N^{\rm th}$ root of unity.  
The contour integral version of the sum is
\begin{equation}
\oint \frac{1}{\alpha - \frac{1}{2}\left(z+ 1/z \right)} 
\frac{Nz^{N-1}}{z^{N}-1} dz
        \label{contint}
\end{equation}
The contour is actually a set of $N$ contours, each going 
counter-clockwise about one of the $N$ roots of $z^{N} -1$.  To 
express the temperature dependence of the sum, we replace $\alpha$ by 
$1+t$.  The roots of $\alpha - \frac{1}{2}\left(z+\frac{1}{z} \right)$ 
then lie on the real $z$-axis, inside and outside the unit circle.  If 
we call them $r_{1}$ and $r_{2}$, where $r_{1}$ is the root that lies 
outside of the unit circle, then
\begin{equation}
r_{1} = 1+t + \sqrt{t^{2} +2 t}
        \label{r1val}
\end{equation}
while $r_{2} = 1/r_{1}$.  The result of the deformation of the contour 
is a sum of two terms, one associated with a contour encircling each 
of the the two roots $r_{1}$ and $r_{2}$.  The end result is
\begin{equation}
\frac{2 Nr_{1}}{r_{1}^{2}-1} \frac{r_{1}^{N}+1}{r_{1}^{N}-1}
        \label{endresult}
\end{equation}
When $t$ is small,
\begin{equation}
r_{1} \approx e^{\sqrt{2t}}
        \label{r1app}
\end{equation}
The corrections to the infinite system limit go as $e^{-\sqrt{2t}N}$.

For comparison, we now approximate cosine functions by the first two 
terms in the expansion in their arguments.  The sum in (\ref{sum1b}) 
is replaced by
\begin{equation}
\sum_{n=-\infty}^{\infty}\frac{1}{t+ \frac{2 \pi^{2}}{N^{2}}n^{2}} = 
\frac{1}{2 \pi i} \int_{c}\frac{N^{2}}{2 \pi^{2}} 
\frac{1}{\frac{N^{2}}{2 \pi ^{2}}t + z^{2}}\frac{\pi}{\tan \pi z} dz
        \label{contint1}
\end{equation}
where the contour of integration is a set of contours, each circling 
counterclockwise about the zeros of $\tan \pi z$, which lie on the 
real axis at integer values of $z$.  These contours can be deformed 
into two, one going around the zero of $\frac{N^{2}}{2 \pi^{2}}t + 
z^{2}$ in the top half plane and the other going around the zero of 
that function of $z$ in the bottom half plane.  The evaluation of 
residues leaves us with the final result
\begin{equation}
\frac{N}{\sqrt{2 t}} \frac{1}{\tanh N\sqrt{t/2}}
        \label{finalresult}
\end{equation}
The expressions (\ref{endresult}) and (\ref{finalresult}) have the 
same limiting values at nonzero $t$, $N \rightarrow \infty$ and at 
large $N$ as $t \rightarrow 0$.  In fact, one can rewrite 
(\ref{endresult}) as
\begin{equation}
\frac{2 Nr_{1}}{r_{1}^{2}-1} \frac{1}{\tanh N \sqrt{t/2}} = \frac{ N 
(1+\sqrt{2t})}{\sqrt{2t} + t} \frac{1}{\tanh N \sqrt{t/2}}
        \label{endresult2}
\end{equation}

The exponentially small finite size corrections are, in leading order, 
identical.

Thus, a truncation of the Fourier transform of the interaction 
potential does not influence the asymptotic form of the finite size 
corrections, if there is not also a cutoff in momentum space.  Of 
course, some sort of cutoff is required in two or more dimensions in 
order that the sum with truncated interaction potential does not 
suffer an ultraviolet (large $q$) divergence

\section{The influence of a soft cutoff on the range of the 
interaction}
\label{app:soft}

The long-range interactions and correlations derived in Appendix 
\ref{app:cutoff} are the consequences of both a truncation in the 
$q$-space expansion of the interaction and the existence of a sharp 
cutoff in momentum space.  To see how the ``roundedness'' of the 
cutoff changes the range of interactions, we modify the sum over $q$ 
in the one dimensional space by introducing a cutoff function, having 
the form
\begin{equation}
C(q,Q) = \frac{1}{1+e^{(q^{2}-1)/Q^{2}}}
        \label{cutoff1}
\end{equation}
In the limit $Q \rightarrow 0$, this cutoff approaches a step 
function.  For finite $Q$, the gradual nature of the cutoff causes the 
interaction to be intrinsically short-ranged.  The real-space version 
of the truncated interaction is, in one dimension,
\begin{equation}
V(x)= \int_{-\infty}^{\infty} \frac{\left( a+bq^{2}\right) 
e^{iqx}}{1+e^{(q^{2}-1)/Q^{2}}} dq
        \label{cutoff2}
\end{equation}
The behavior of the interaction can be extracted by distorting the 
contour of integration in (\ref{cutoff2}) so that it encloses the 
poles of the cutoff function, which are at the locations in the 
complex $q$ plane at which
\begin{equation}
q=\left(1 \pm i Q^{2} \left(n + \frac{1}{2}\right) \pi\right)^{1/2} 
\approx 1 \pm i \pi Q^{2} \frac{2n +1}{4}
        \label{root}
\end{equation}
where the approximate equality holds if $Q$ is small and the arbitrary 
integer $n$ is not too large.  The residue at such poles has the 
$x$-dependence $e^{ - Q^{2} \frac{2n+1}{4} \pi x}$.  This exponential 
decay of interactions, and hence of ``intrinsic'' correlations, will 
not give rise to effects interfering with the finite size corrections 
that go as $e^{-L/ \xi}$.  This can be seen by, first, repeating the 
analysis of Appendix \ref{app:cutoff}.  Alternatively, one can look at 
the sum
\begin{equation}
\sum_{q}\frac{1}{r+q^{2}}\frac{1}{1+e^{(q^{2}-1)/Q^{2}}}
        \label{softsum}
\end{equation}
where the allowed values of $q$ are
\begin{equation}
q=\frac{2 \pi n}{L}
        \label{qallowed}
\end{equation}
Again, $L$ is the size of the system in units of the lattice spacing 
between spins.  The sum in (\ref{softsum}) is evaluated with the use 
of the Poisson sum formula:
\begin{equation}
\sum_{q} f(q) = \sum_{m=-\infty}^{\infty} \int dq f(q) e^{imqL}dq
        \label{Poisson1}
\end{equation}
The $m=0$ contribution to the sum in question is just the infinite 
system limit of the sum in the equation of state.  Finite size 
corrections arise from the $m \neq 0$ terms in that sum.  Such terms 
have the form
\begin{equation}
\int_{-\infty}^{\infty} 
\frac{1}{r+q^{2}}\frac{e^{imqL}}{1+e^{(q^{2}-1)/Q^{2}}}dq
        \label{Poisson2}
\end{equation}
We evaluate the integral in the same way as we did in the case of the 
interaction, except that here there is an additional pole in the 
complex $q$-plane, at the root of the denominator in $1/(r+q^{2})$.  
The residue at this pole yields a contribution going as 
$e^{-m\sqrt{r}L} = e^{-mL/\xi}$.  The residues of the poles of the 
cutoff function give rise to terms going as $e^{ - mQ^{2} 
\frac{2n+1}{4} \pi L}$, when $Q$ is small and $n$ is not too large.  
As the critical point is approached, $r$ becomes small, the 
contribution going as $e^{-\sqrt{r}L}$ dominates all others, and 
finite size scaling in the expected form is recovered.

\section{Alternative approach to the analysis of the equation of 
state}
\label{app:alternative}

\begin{figure}
\centerline{\epsfig{file=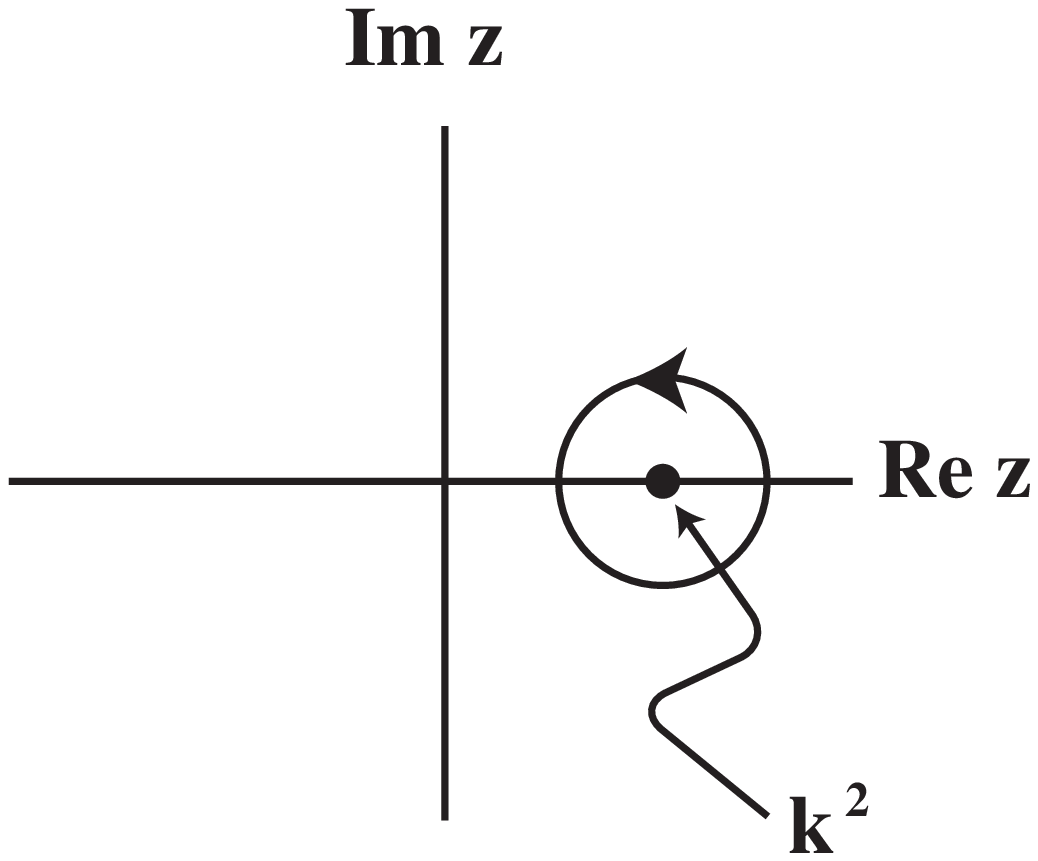,height=2in}}
\caption{The contour utilized in the contour integration identity 
(\ref{oint1}) for the summand in the correction to the equation of 
state.}
\label{fig:contour1}
\end{figure}

\begin{figure}
\centerline{\epsfig{file=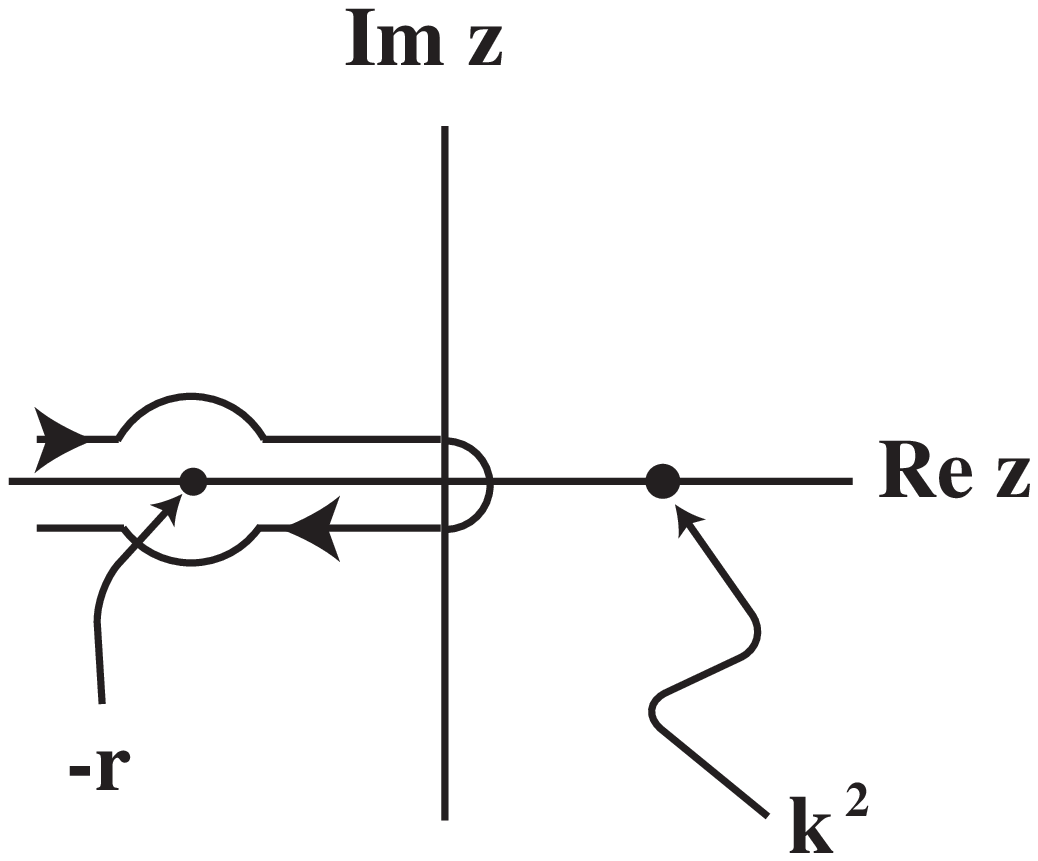,height=2in}}
\caption{The deformation of the contour in Figure \ref{fig:contour1} 
that leads to the new expression (\ref{oint2}) for the summand in the 
correction to the equation of state.}
\label{fig:contour2}
\end{figure}

The principal influence of the subleading long-range contribution to 
the spin-spin interaction is obtained by expanding to first order in 
that interaction.  In the case of the equation of state, the 
correction term, exhibited on the first line of Eq.  (\ref{eosa}), is
\begin{equation}
\sum_{\vec{k}}\frac{k^{\sigma}}{\left(r+k^{2}\right)^{2}}
        \label{sum1}
\end{equation}
The exponent $\sigma$ is greater than one.  We start by making use of 
the contour integration identity
\begin{equation}
\frac{k^{\sigma}}{\left(k^{2}+r\right)} = \frac{1}{2 \pi i} \oint 
\frac{z^{\sigma/2}}{\left(z+r\right)^{2}}\frac{1}{z-k^{2}}dz
        \label{oint1}
\end{equation}
where the integral is around the contour that encircles the pole of 
the integrand at $z=k^{2}$.  Figure \ref{fig:contour1} shows the 
contour over which the integration is performed.  The integrand in 
(\ref{oint1}) has, in addition to the abovementioned pole, a branch 
point at $z=0$, which results from the term $z^{\sigma/2}$, assuming 
that the quantity $\sigma/2$ is not an integer.  There is also a 
double pole at $z=-r$.  The integral in (\ref{oint1}) is evaluated by 
deforming the contour so that it surrounds the branch cut from $z=0$ 
to $z=-\infty$, with a special accommodation at the pole at $z=-r$.  
Figure \ref{fig:contour2} is a picture of the deformed contour.  The 
integral over this new contour has the form
\begin{equation}
-\frac{\sin \sigma \pi /2}{ 
\pi}\int_{0}^{\infty}\frac{z^{\sigma/2}}{(r-z)^{2}}\frac{1}{z+k^{2}} 
dz +\cos \left(\sigma \pi / 2\right) \left.  
\frac{d}{dz}\left[\frac{z^{\sigma/2}}{z+k^{2}}\right]\right|_{z=r}
\label{oint2}
\end{equation}
In this expression, the integration variable $z$ has been replaced by 
$-z$.  The integration in the first term in (\ref{oint2}) is 
understood to be in the form of a principal parts integral.  Such an 
integration combined with the second term has the effect of removing 
the non-integrable singularity at $z=r$.  A new form for (\ref{oint2}) 
is obtained with the use of integration by parts.  The end result is 
an integral of the form
\begin{equation}
 \cos \pi \sigma /2 
\frac{d}{dr}\left[\frac{r^{\sigma/2}}{r+k^{2}}\right] + \frac{\sin 
\pi \sigma /2}{\pi}\int_{0}^{\infty} \frac{1}{r-z} \frac{d}{dz}\left[ 
\frac{z^{\sigma/2}}{z+k^{2}}\right] dz
        \label{oint3}
\end{equation}
The integral in (\ref{oint3}) is understood as a principal parts 
integral.

The next step is to perform the sum over $k$.  This sum is fairly 
straightforward, in that it is the one encountered in studies of 
finite systems with short range interactions \cite{RGJ,SR}.  That has 
been done previously.  In $d$ dimensions, the result of the summation 
is given by
\begin{eqnarray}
\sum_{\vec{k}}\frac{1}{z+k^{2}} &=& \frac{L^{2}}{4 \pi} 
\int_{1}^{\infty} \left[ e^{-zL^{2}t/ 4 \pi} + e^{-zL^{2}/4 \pi t} 
\right] \left[ Q(t)^{d} -1 \right] dt \nonumber \\
&&- \frac{L^{2}}{4 \pi} \int_{1}^{\infty}e^{-zL^{2}t/4 \pi} t^{-d/2}dt 
+\frac{1}{z}e^{-4 \pi z L^{2}} + \frac{L^{d}}{\left(2 \pi \right)^{d}} 
\int \frac{d^{d}k}{k^{2}+z} \nonumber \\
\label{sumval}
\end{eqnarray}
where
\begin{equation}
Q(t) = \sum_{n=-\infty}^{\infty}e^{- \pi n^{2}t}
        \label{Q(t)}
\end{equation}
If we extract the infinite system term---the last term on the 
right---from the right hand side of Eq.  (\ref{sumval}), we are left 
with a function that has the general form
\begin{equation}
L^{2}g(zL^{2})
        \label{genform}
\end{equation}
In addition, this function decays exponentially with large values of 
$zL^{2}$.  In fact, it goes as $e^{-zL^{2}/4 \pi}$.  For small values 
of $z$, the function is dominated by the term $1/zL^{2}$.

\begin{figure}
\centerline{\epsfig{file=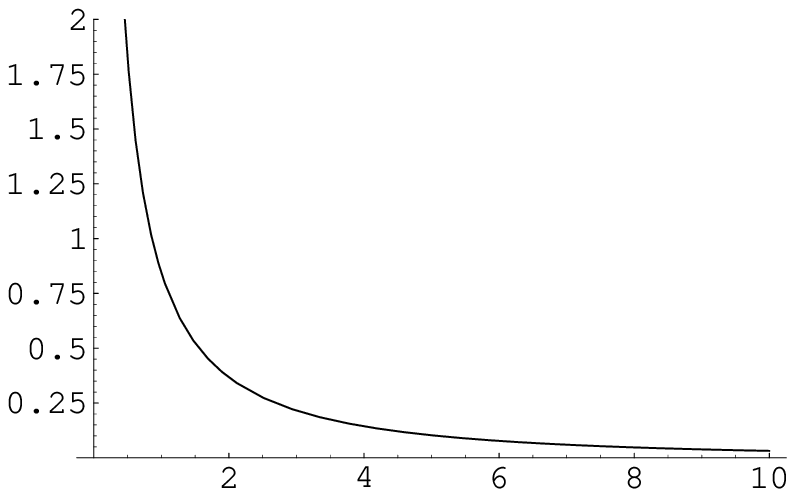,height=2in}}
\caption{The sum $\sum_{\vec{k}}1/(z+k^{2})$, as a function of 
$zL^{2}$, divided by $L^{2}$.  The infinite system limit to the sum 
has been removed.}
\label{fig:graph1}
\end{figure}

Figure \ref{fig:graph1} is a graph of the sum, with the infinite 
system term removed.  Given the general form of this function, we can 
write the expression for the correction to the leading order 
contribution to the equation of state.  It has the form
\begin{equation}
\cos \left( \sigma \pi /2 \right) \frac{d}{dr} \left[ r^{\sigma /2} 
L^{2} g(rL^{2}) \right] + \frac{\sin \sigma \pi /2}{ 
\pi}\int_{0}^{\infty}\frac{1}{r-z}\frac{d}{dz} \left[z^{\sigma 
/2}L^{2}g(zL^{2})\right] dz
        \label{new}
\end{equation}

The expression above for the correction to the equation of state can 
be shown to have the expected properties in various regimes. For 
example, if $rL^{2}$ is large, then the leading order behavior arises 
from the second term in (\ref{new}). If we rescale the variable of 
integration by making the replacement $z \rightarrow z/L^{2}$, the 
integral in (\ref{new}) becomes
\begin{equation}
\frac{\sin \sigma 
\pi}{\pi}L^{4-\sigma}\int_{0}^{\infty}\frac{1}{rL^{2}-z} 
\frac{d}{dz}\left[z^{\sigma/2}g(z) \right] dz \equiv 
L^{4-\sigma}{\cal G}(rL^{2})
        \label{trans1}
\end{equation}
At large values of $rL^{2}$, the integrand can be expanded in inverse 
powers of that combination. The lowest order term, going as 
$1/rL^{2}$, can be shown to integrate to zero. The next order term is
\begin{equation}
\frac{\sin \sigma \pi}{\pi}\frac{L^{-\sigma}}{r^{2}}\int_{0}^{\infty}z 
\frac{d}{dz}\left[z^{\sigma/2}g(z) \right] dz =-\frac{\sin \sigma 
\pi}{\pi}\frac{L^{-\sigma}}{r^{2}}\int_{0}^{\infty} 
\left[z^{\sigma/2}g(z) \right] dz
        \label{limit1}
\end{equation}

Finally, use of the identity
\begin{equation}
\int_{0}^{1}\frac{x^{-p}-x^{p-1}}{1-x}dx = - \pi \cot p 
\pi \ \ \ \ \  0<p<1
        \label{cancelident}
\end{equation}
allows us to show that the expression in Eq.  (\ref{new}) for the 
correction to the equation-of-state sum does not give rise to any 
terms going as a fractional power of $r$ in the limit $r \rightarrow 
0$.  This is consistent with the expectations one has for the limiting 
behavior of a finite system, and with the analysis in Sections 
\ref{sec:eos} and \ref{sec:chi}.

\section{The equation of state in general}
\label{app:general}

While an analysis sufficient for our purposes can be carried out by 
focusing entirely on the first order effect of the long-range 
component of the interaction to, say, the equation of state, it is 
also possible to write down an expression for the entire equation of 
state for the system with a subleading long-range interaction.  We 
start with the identity
\begin{equation}
\frac{1}{r+ak^{2}+bk^{\sigma}+ck^{4}} = \frac{1}{2\pi i} \oint 
\frac{1}{r+az+b z^{\sigma/2}+cz^{2}}\frac{1}{z-k^{2}}dz
        \label{int1}
\end{equation}
Where the closed integration contour is about the pole in the 
integrand at $z=k^{2}$.  See Figure \ref{fig:contour1}. 

\begin{figure}
\centerline{\epsfig{file=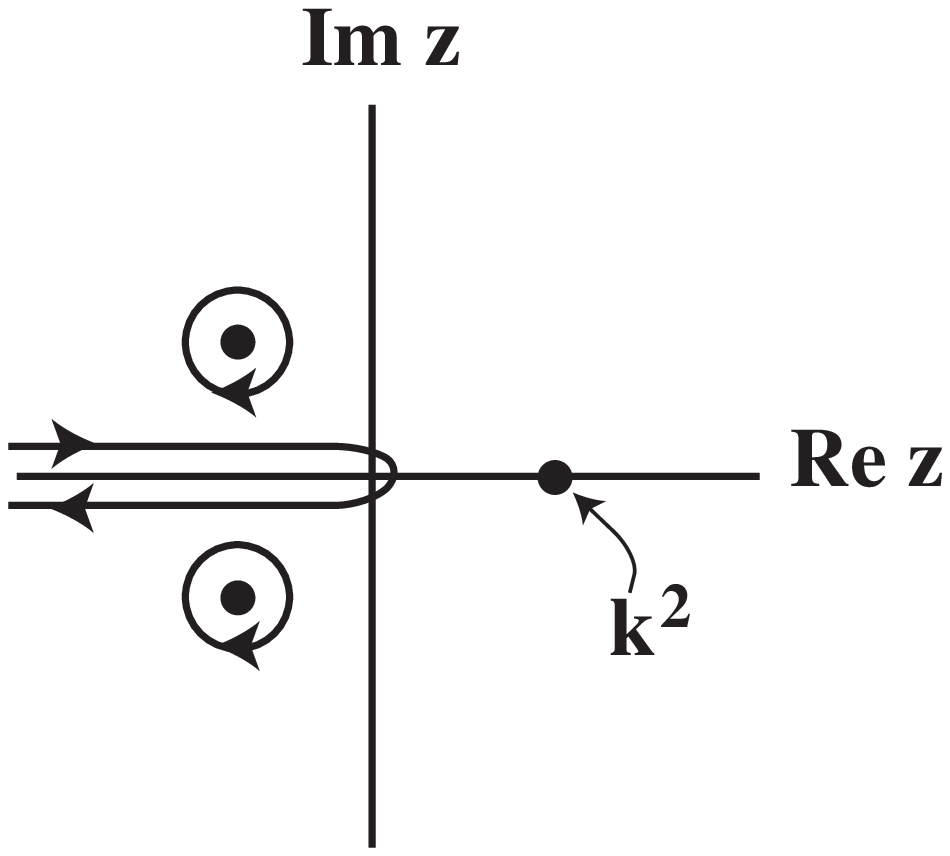,height=1.5in}}
\caption{The distortion of the contour displayed in Figure 
\ref{fig:contour1} that leads to the expression (\ref{lhs1}) for the 
equation of state}
\label{fig:cont2}
\end{figure}

The next step is to distort the contour so that it wraps around the 
two poles of $1/(r+az+bz^{\sigma/2}+cz^{2})$, and around the branch 
cut on the negative $z$-axis.  These poles exist for the range of 
$\sigma$ considered here: $4>p \sigma >2$.  The distorted contour is 
as depicted in Figure \ref{fig:cont2}.  In the case of the two roots 
in the right half of the $z$-plane, there is a contribution from the 
residue proportional to the inverse of the derivative with respect to 
$z$ of $r+az+bz^{\sigma/2}+cz^{2}$.  For the contour integral, because 
the direction of integration is different on the two sides of the 
branch cut.  We take the difference between $1/(r-az+b(ze^{\pm i 
\pi})^{\sigma/2}+cz^{2}) = 1/(r-az +cz^{2}+b z^{\sigma /2} \cos \sigma 
\pi /2 \pm i b z^{\sigma /2} \sin \sigma \pi /2)$.

\begin{eqnarray}
\sum_{\vec{k}}\frac{1}{r+ak^{2} +bk^{\sigma}+ck^{4}} &=& \frac{1}{r} 
+\sum_{\vec{k}\neq 0}\left\{ 2 {\Re} \left[\frac{1}{a+b 
(\sigma/2) z_{0}^{\sigma/2 -1}+2cz_{0}}\frac{1}{k^{2}-z_{0}}\right] 
\right.  \nonumber \\
&& \left.  +\frac{1}{\pi} \int_{0}^{\infty}\frac{b z^{\sigma/2} \sin 
\sigma \pi /2}{\left(r-a z + cz^{2} +b z^{\sigma /2} \cos \sigma \pi /2 
\right)^{2} + b^{2} z^{\sigma} \sin^{2} \sigma \pi /2} 
\frac{1}{k^{2}+z} dz \right\} \nonumber \\
        \label{lhs1}
\end{eqnarray}
In Eq.  (\ref{lhs1}), the quantity $z_{0}$ is the solution of the 
equation
\begin{equation}
r+a z+b z^{\sigma /2}+cz^{2} =1
        \label{eqforz0}
\end{equation}
There are actually two solutions to (\ref{eqforz0}).  When $\sigma$ 
\raisebox{-.6ex}{$\stackrel{>}{\sim}$} 2, they lie just above and 
below the negative real $z$-axis.  In fact, as $ r \rightarrow 0$, $ 
z_{0} \rightarrow -r/\sqrt{a} \pm i \delta$, with $\delta \ll r$.  We will 
assume that the $z_{0}$ that enters into Eq.  (\ref{lhs1}) is the 
solution lying in the upper half of the complex $z$-plane.  The sum 
over $\vec{k}$ in (\ref{lhs1}) is carried out in the standard way, the 
result being given by Eq.  (\ref{sumval}).

While the resulting expression for the sum over $\vec{k}$ in the 
equation of state is correct to all orders in the amplitude of the 
singular contribution to the interaction, the expression presents 
significantly greater challenges to the theorist, and all important 
effects are recovered from the first order term in the expansion with 
respect to the long-range interaction.

\section{Estimate of the leading $\Lambda$ dependence of the
  finite-size term}
\label{app:Lambdep}
We are interested in determining the leading $\Lambda$ dependence of
the sum
\begin{equation}
W_{d}^{\sigma}(r,b,L,c|\Lambda)=\frac{1}{L^d}\sum_{\bf 
q}\frac{1}{r+q^2+bq^\sigma+c q^4}.
\end{equation}
We will assume a sharp cutoff, i.e $\bf q$ is a vector with components 
$q_j=2\pi m_j/L$, $m_j=\pm 1,\pm2, \cdots$, $j=1,2, \cdots, d $, in 
the range $-\Lambda\le q_j<\Lambda$.  The leading $\Lambda$-dependence 
of the sum arises from terms with large $q$. One can, therefore, omit 
the $q^2$ and $q^\sigma$ contributions.  Let us denote the leading 
$\Lambda$-dependent term of $W_{d}^{\sigma}(r,b,c,L|\Lambda)$ by $\Delta 
W_{d}^{\sigma}(r,b,c,L|\Lambda) $.  Then
\begin{equation}
\Delta W_{d}^{\sigma}(r,b,c,L|\Lambda)=
\frac{1}{(2\pi)^4 b y L^{d-4}}\sum_{m_1=-M}^{M-1}\cdots
\sum_{m_d=-M}^{M-1}
\frac{1}{1+ (m^2/\sqrt{y})^{2}},
\end{equation}
where $y=(r/b)(L/2 \pi)^4$ and $m=(m_1^2 + \cdots +
m_d^2)^{1/2}$. 
Using the identity \cite{BDT00}
\begin{equation}
\frac{1}{1+z^\alpha}=\int_0^\infty dt \  e^{-zt}t^{\alpha-1}
E_{\alpha,\alpha}(-t^\alpha),  
\end{equation}
where $E_{\alpha,\alpha}(x)$ are the Mittag-Leffler functions, the
above expression can be rewritten in the form 
\begin{equation}
\Delta W_{d}^{\sigma}(r,b,c,L|\Lambda)=
\frac{1}{(2\pi)^4 b y L^{d-2}} \int_0^{\infty}dx
x
E_{2,2}(-x^{2})\left[\sum_{m=-M}^{M-1}
e^{-x m^2/\sqrt{y}}\right]^d. 
\end{equation}
Taking into account that $E_{2,2}(z)=\sinh z/\sqrt{z}$ \cite{BDT00}
with the help of Eqs. (\ref{cd14}) and (\ref{cd15}) we obtain 
\begin{equation}
\Delta W_{d}^{\sigma}(r,b,c,L|\Lambda) = -\frac{d}{48 \pi^{(9-d)/2}}  
M L^{d-4}\int_0^\infty dx 
x^{(5-d)/2} \frac{\sin(x\sqrt{y})}{x\sqrt{y}}
\left[{\rm erf}(M\sqrt{x})\right]^{d-1}
e^{-M^2 x}.
\end{equation}
Since $M \gg 1$ in the above equation, the only significant 
contributions are those stemming from small $x$.  After taking into 
account that $M^2\gg y^{2/\sigma}$, and, so, one has  
$\lim_{x\rightarrow 0} \sin(x\sqrt{y})/(x\sqrt{y})=1$, we are led to
\begin{equation}
  \Delta
  W_{d}^{\sigma}(r,b,c,L|\Lambda)=O(L^{-(d-4)}M^{d-6})=
O(M^{-2}\Lambda^{d-4}).
\end{equation}
The contribution in the equation of state that does not depend on 
$\Lambda$, but is due to the long-range character of the interaction is of 
order $O(L^{4-d-\sigma})$.  Elementary checks reveal that when 
$d<4$, $d+\sigma\le 6$, $L\gg 1$ and $\Lambda \gg 1$ one has 
$L^{6-d-\sigma} \gg \Lambda^{d-6}$, whence,
\begin{equation}
  \Delta
  W_{d}^{\sigma}(r,b,c,L|\Lambda)\ll L^{4-d-\sigma}.
\end{equation}
Thus, up to the order at which the results are presented in this 
article, those results will not be influenced by the (nonuniversal) 
$\Lambda$-dependent corrections.

\section{Derivation of the leading asymptotics of the bulk term 
$W_{d}^{\sigma}(r,b|\Lambda) $}
\label{thebulkder}
As we are interested only in the contribution stemming from small 
$q$, the expansion below is justified and one obtains
\begin{eqnarray}
W_{d}^{\sigma}(r,b,c|\Lambda) &\equiv& 
\frac{1}{(2\pi)^d}\int_{-\Lambda}^{\Lambda}d^dq \cdots 
\int_{-\Lambda}^{\Lambda} \frac{1}{r+q^2+bq^\sigma+ c q^4} \nonumber \\
&=& W_{d}^{\sigma}(0,b,c|\Lambda)- 
r\frac{1}{(2\pi)^d}\int_{-\Lambda}^{\Lambda}d^dq \cdots 
\int_{-\Lambda}^{\Lambda} \frac{1}{q^2(r+q^2)} \nonumber \\
& & +br\frac{1}{(2\pi)^d}\int_{-\Lambda}^{\Lambda}d^dq \cdots 
\int_{-\Lambda}^{\Lambda} \frac{q^{\sigma-2}}{(r+q^2)^2} \nonumber \\
& & +br\frac{1}{(2\pi)^d}\int_{-\Lambda}^{\Lambda}d^dq \cdots 
\int_{-\Lambda}^{\Lambda} \frac{q^{\sigma-4}}{r+q^2} \nonumber \\
& & -rb^2\frac{1}{(2\pi)^d}\int_{-\Lambda}^{\Lambda}d^dq \cdots 
\int_{-\Lambda}^{\Lambda} \frac{q^{2(\sigma-2)}}{(r+q^2)^2} + \cdots.
\label{bulkas}
\end{eqnarray}
The nonanalitycity in the behavior of all this function arises 
entirely from small-$q$ contributions in the integrals.  We 
sphericalize the region of integration and extend the limits of 
integration from zero to infinity.  If the corresponding integral 
diverges after such a procedure, we first differentiate the requisite 
number of times with respect to $r$, perform a replacement of the 
limits of integration in the first derivative that does not diverge, 
calculate the leading $r$ behavior, and, finally, integrate the 
required number of times with respect to $r$.  Performing this 
procedure we obtain ($d+\sigma<6$, $2<d<4$, $2<\sigma<4$)
\begin{eqnarray}
W_{d}^{\sigma}(r,b,c|\Lambda) &=& W_{d}^{\sigma}(0,b,c|\Lambda)\nonumber 
\\
& & +\frac{\pi}{(4\pi)^{d/2}\Gamma(d/2)\sin(\pi d/2)} 
r^{d/2-1}\nonumber \\
& & +b \frac{\pi(d+\sigma-4)}{2(4\pi)^{d/2}\Gamma(d/2)\sin(\pi( 
d+\sigma)/2)} r^{d/2-1+(\sigma-2)/2}\nonumber \\
& &+ b \frac{\pi}{(4\pi)^{d/2}\Gamma(d/2)\sin(\pi( 
d+\sigma)/2)} r^{d/2-1+(\sigma-2)/2}\nonumber \\
& &+O(r^{(d+2(\sigma-2))/2-1},r),
\end{eqnarray}
wherefrom we are able to obtain the result given in the main body of 
the article.

In a similar way, one can treat the case $d+\sigma=6$ with $2<d<4$, 
$2<\sigma<4$.

\section{Derivations of the asymptotic form of the non-leading 
long-range correction term}
\label{lras}

In this appendix, we provide details of the calculation leading to the 
asymtotic form of the expression $\left(1+x \partial  / \partial x 
\right) I^{p}_{\rm scaling}(x,d)$, for various ranges of the variable 
$x$. We begin with the result
\begin{eqnarray}
I_{{\rm scaling}}^p(x,d) &= & \int_{0}^{\infty} e^{-xt}t^{-p} 
\gamma^{*}(-p,-xt) \left\{ \sum_{\vec{k}}e^{-4 \pi^{2}k^{2}t} - 
\left( 4 \pi t \right)^{-d/2} -1 \right\} dt
\nonumber
\\
& = & \left( 4 \pi \right)^{-d/2} \sum_{\vec{k} \neq 0} 
\int_{0}^{\infty} e^{-xt}t^{-(d/2 +p)} \gamma^{*}(-p,-xt) e^{-k^{2}/4t} 
dt
\label{h1}
\end{eqnarray}
where use has been made of the Poisson sum formula and the fact that
\begin{equation}
\int_{0}^{\infty} e^{-xt}t^{-p} \gamma^{*}(-p,-xt) dt =0
        \label{h2}
\end{equation}
The proof of this last statement is relatively straightforward.  Making 
use of the series representation of the function $\gamma^{*}$ one 
obtains
\begin{eqnarray}
\int_{0}^{\infty}e^{-xt} t^{-p} \gamma^{*}(-p, -xt) dt &=& x^{-1+p} 
\int_{0}^{\infty} e^{-t}t^{-p} \gamma^{*} (-p, -t) dt \nonumber \\
&=& x^{-1+p} \frac{1}{\Gamma(-p)} \sum_{n=0}^{\infty} 
\int_{0}^{\infty}\frac{t^{n-p}}{(n-p) n!} dt \nonumber \\
&=& \frac{x^{-1+p}}{\Gamma(-p)} \sum_{n=1}^{\infty} 
\frac{\Gamma(1+n-p)}{(n-p)n!} \nonumber \\
&=& 0
\label{h3}
\end{eqnarray}
The last equality in (\ref{h3}) holds because
\begin{equation}
(1-x)^{p} = 
\frac{1}{\Gamma(-p)}\sum_{n=0}^{\infty}\frac{\Gamma(1+n-p)}{(n-p)n!}x^{n}
        \label{h4}
\end{equation}

From Eq. (\ref{h1}) it is clear that, if $x \rightarrow \infty$ and 
$xt = O(1)$, then the contributions arising from $t \sim 1/x$ will be 
exponentially small in $x$ because of the term going as $e^{-k^{2}/4t}$. 
Therefore, the leading order contributions in the regime $x 
\rightarrow \infty$ will be generated by the asymptotics of 
$\gamma^{*}(-p, xt)$ in the regime $xt \gg 1$. Making use of the 
identity \cite{AS}
\begin{equation}
\frac{\partial^{n}}{\partial x^{n}}\left[ e^{x}x^{a} \gamma^{*}(a,x) 
\right] = e^{x}x^{a-n} \gamma^{*}(a-n,x)
        \label{h5}
\end{equation}
One can write $(1+ x \partial /\partial x) I^{p}_{\rm scaling}(x,d)$ 
in the following form
\begin{eqnarray}
\left(1+x \frac{\partial}{\partial x}\right) I^{p}_{\rm 
scaling}(x,d) &=& \left( 4 \pi \right)^{-d/2} \sum_{\vec{k}\neq 
0}\int_{0}^{\infty}e^{-xt}t^{d/2 + p} e^{-k^{2}/4t} \nonumber \\ 
&&\times \left[(p+1) 
\gamma^{*}(-p,-xt) + \gamma^{*}(-p-1,-xt) \right] dt
        \label{h6}
\end{eqnarray}
We now need the asymptotics of the function $\gamma^{*}(-p,-x)$ for 
$x \rightarrow \infty$. The leading order behavior of the function in 
this limit follows form the relationship
\begin{equation}
\gamma^{*}(-a,-x) = \frac{1}{\Gamma(1-a)}M(-a,1-a,x)
        \label{h7}
\end{equation}
where $M(a,b,z)$ is the Kummer function, and the corresponding 
asymptotic behavior of that function is \cite{AS}
\begin{equation}
M(a,b,z) = \frac{\Gamma(b)}{\Gamma(a)} e^{z}z^{a-b} \left[ 
1+O(|z|^{-1}) \right], \ \ \ \ \ \Re z > 0, |z| \rightarrow \infty
        \label{h8}
\end{equation}
One then obtains
\begin{equation}
\gamma^{*}(-a,-x) \simeq \frac{1}{\Gamma(-a)}e^{x}x^{-1} 
\left(1+\frac{C}{x}\right), \ \ \ \ a<1, \ x \gg 1
        \label{h9}
\end{equation}
where $C$ is a constant. In order to determine $C$ we make use of the 
identity \cite{AS}
\begin{equation}
\gamma^{*} (a,-x) =-x\gamma^{*}(a+1,-x) + 
\frac{e^{x}}{\Gamma(a+1)}
        \label{h9a}
\end{equation}
Making use of this equation and (\ref{h9}) we find that $C=a$, i.e.
\begin{equation}
\gamma^{*}(-a,-x) \simeq \frac{1}{\Gamma(-a)}e^{x}x^{-1} 
\left(1+\frac{a}{x}\right), \ \ \ \ a<1, \ x \gg 1 .
        \label{h9b}
\end{equation}
Inserting this result into (\ref{h6}), we obtain
\begin{equation}
\left(1+ x \frac{\partial}{\partial x} \right) I^{p}_{\rm 
scaling}(x,d) \simeq C_{p}x^{-2}
        \label{h10}
\end{equation}
where
\begin{equation}
C_{p} = - \frac{1+p}{\Gamma(-p) \left( 4 \pi \right)^{d/2}} 
\sum_{\vec{k} \neq 0} \int_{0}^{\infty} t^{-(2+p+d/2)}e^{-k^{2}/4t} 
dt .
        \label{h11}
\end{equation}
The above holds when $0<p<1$. It is easy to show that $C_{p} >0$. 
If the integral in (\ref{h11}) is performed, $C_{p}$ is recast in the 
form
\begin{equation}
C_{p} = -\frac{(1+p) 
4^{1+p}}{\pi^{d/2}}\frac{\Gamma(1+p+d/2)}{\Gamma(-p)}\sum_{\vec{k} 
\neq 0}\frac{1}{k^{d+2(p+1)}}.
        \label{h12}
\end{equation}
In terms of Madelung type constants $C(d\ |a)$ one can rewrite $C_p$ in the form 
\cite{CT2}
\begin{equation}
C_p=-\frac{1+p}{\Gamma(-p)}(2\pi)^{2(p+1)} C(d \ \left | \right.
\frac{d}{2}+p+1),  
\end{equation}
or, equivalently, in terms of Epstein zeta function $\cal Z$ this constant is
\begin{equation}
  C_p=-\frac{(1+p)4^{1+p}}{\pi^{d/2}}\frac{\Gamma(1+p+d/2)}{\Gamma(-p)}
{\cal Z} \left |\begin{array}{c} 0\\0
\end{array} \right| (d,\frac{d}{2}+p+1). 
\end{equation}
For the case $d=\sigma=3$ (then $p=1/2$) the numerical 
evaluation gives $C_{1/2}=10.216$.

When the parameter $p$ is equal to zero, appropriate to the case of 
short-range interactions, the asymptotic form of interest is of the 
function
\begin{eqnarray}
I_{s}^{0}(x) &\equiv & \int_{0}^{\infty} dt e^{-xt} \left\{ 
\sum_{\vec{k}} e^{- 4 \pi^{2} k^{2} t}- \left(4 \pi t \right)^{-d/2} 
-1 \right\} \nonumber \\
& = & \left( 4 \pi \right) ^{-d/2} \sum_{\vec{k} \neq 0} 
\int_{0}^{\infty}  t^{-d/2} e^{-xt -k^{2}/4t} dt  -1/x \nonumber \\
& = & -\frac{1}{x} + x^{d/4 -1/2} \pi^{-d/2} \sum_{\vec{k} \neq 0} 
k^{-(d-2)/2}K_{d/2 -1}\left( k \sqrt{x}\right)
        \label{h13}
\end{eqnarray}
We now make use of the asymptotic form of the modified Bessel Function:
\begin{equation}
K_{\nu}(x) \simeq \sqrt{\frac{\pi}{2x}}e^{-x}, \ \ \ \ \ x \gg 1.
        \label{h14}
\end{equation}
We immediately find that, for $x \gg 1$,
\begin{equation}
I_{s}^{0}(x) = -\frac{1}{x} + \frac{d 
\sqrt{2}}{\pi^{(d-1)/2}}x^{(d-3)/4}e^{-\sqrt{x}}
        \label{h15}
\end{equation}

Let us now derive the asymptotic form of $(1+x \partial / \partial 
x)I^{p}_{\rm scaling}(x,d)$ for $x \ll 1$. To that end one needs only to
note that 
\begin{equation}
e^{x} \gamma^{*}(a,x) \rightarrow \frac{1}{\Gamma(a+1)}
        \label{h16}
\end{equation}
when $x \rightarrow 0$. Then, as $x \rightarrow 0$,
\begin{eqnarray}
I^{p}_{\rm scaling}(x,d) &\simeq& I^{p}_{\rm scaling}(0,d) \nonumber \\
&=& \frac{1}{\Gamma(1-p)} \int_{0}^{\infty} t^{-p} \left\{ 
\sum_{\vec{k} \neq 0} e^{-4 \pi^{2}k^{2}t} - \left(4 \pi t 
\right)^{-d/2} \right\} dt \nonumber \\
& = & \frac{1}{\Gamma(1-p)} \int_{0}^{\infty} t^{-p} \left[ \left(4 
\pi t \right)^{-d/2} \sum_{\vec{k} \neq 0}e^{-k^{2}/4t} -1 \right] dt
\label{h17}
\end{eqnarray}
The integrands on the right hand side of Eq.  (\ref{h17}) are 
well-defined for $p <1$ and $2 < d < 4$, both at the lower and upper 
bounds of integration. We will denote the above constant by $D_{p}$. 
Then,
\begin{equation}
\left(1+x \frac{\partial}{\partial x} \right)I^{p}_{\rm 
scaling}(x,d) \simeq D_{p}, \ \ \ \ \ 0<p<1, \ \ \ \ \ x \rightarrow 0.
        \label{h18}
\end{equation} 
When $p=0$, it is straightforward to show that
\begin{equation}
I^{0}_{\rm scaling}(x,d) \simeq I_{\rm scaling}^{0}(0,d) = D_{0}.
        \label{h19}
\end{equation}

It is easy to show that
\begin{equation}
D_p=\frac{1}{\Gamma(1-p)}(2\pi)^{2(p+1)}C(d\ |\frac{d}{2}+p-1).  
\end{equation}
Since the Madelung constants $C(d|a)$ are negative for $d/2>a$, we
obtain that $D_p<0$ (for $p<1$). The numerical evaluation for
$d=\sigma=3$ (then $p=1/2$) gives $D_{1/2}=-0.452$ while $D_0=-0.226$
(which is consistent with $C(3|1)=-5.029$ and $C(3|1/2)=-8.914$, respectively).

%
%

\end{document}